\documentclass{CUP-JNL-DTM}%

\usepackage{graphicx}
\usepackage{multicol,multirow}
\usepackage{amsmath,amssymb,amsfonts}
\usepackage{mathrsfs}
\usepackage{amsthm}
\usepackage{rotating}
\usepackage{appendix}
\usepackage[numbers]{natbib}
\usepackage{ifpdf}
\usepackage[T1]{fontenc}
\usepackage{newtxtext}
\usepackage{newtxmath}
\usepackage{textcomp}
\usepackage{xcolor}
\usepackage{lipsum}
\usepackage[colorlinks,allcolors=blue]{hyperref}

\theoremstyle{definition}

\numberwithin{equation}{section}

\newcommand{\pder}[2]{\frac{\partial #1}{\partial #2}}

\jname{International Journal of Multiphase Flow}
\jyear{2023}

\begin{document}

\begin{Frontmatter}

\title[Article Title]{Planar bubble plumes from an array of nozzles: Measurements and modelling }

\author[1]{Simon Beelen}
\author[1]{Dominik Krug}

\authormark{Beelen \& Krug}

\address[1]{\orgdiv{Physics of Fluids group}, \orgname{University of Twente}, \orgaddress{\city{Enschede}, \postcode{7522NB},  \country{Netherlands}}}


\abstract{Bubble curtains are widely used for sound mitigation during offshore pile driving to protect marine life. However, the lack of well validated hydrodynamic models is a major factor in the inability to predict the sound attenuation of a bubble curtain \emph{a priori}. We present a new dataset resulting from bubble curtain measurements carried out in a 10m deep and 31m wide freshwater tank. The data describe the evolution of the void fraction profile and the bubble size distribution along the height of the bubble curtain. On this basis, a new relationship is developed for the dependence of the entrainment parameter of the bubble curtain on the air flowrate. In addition, we have extended a recently developed integral model for round bubble plumes to seamlessly capture the transition from initially individual round plumes to a planar plume after their merger. With additional modifications to the entrainment relation, the effective slip velocity and the initial condition for the bubble size distribution, the new model is found to be in good agreement with the data. In particular, the bubble size distribution sufficiently distant from the source is found to be independent of the gas flowrate, both in the data and in the model.}

\end{Frontmatter}

\section[Introduction]{Introduction}
Recently, the interest in bubble curtains has increased due to their use in sound mitigation of off-shore pile driving activities. As part of the drive for sustainable energy, many off-shore wind farms are being constructed or planned; e.g. in \citet{EMODNet} an overview of European wind farm projects and their status can be found. Due to the disturbing effects of pile driving on marine life (see e.g. \citet{dahl2015underwater,popper2022offshore}) the European Committee initiated the Marine Strategy Framework Directive (MSFD). This overarching initiative has led member states to set regulations in line with the MSFD. \citet{juretzek2021turning} provide a comprehensive overview of the current regulations regarding the emitted noise during pile driving. Especially the German implementation has been particularly effective as the sound levels are reduced while the development of wind farms remains economically profitable (\citet{merchant2019underwater}). The German regulation is based on a sound limit measured at $750\,\mathrm{m}$ away from the pile driving site, giving the developer the freedom to reduce the emitted noise by their method of choice. Most frequently bubble curtains are employed for this task (\citet{bellmann2014overview}), however \citet{strietman2018measures} show that operating a bubble curtain is expensive and costs can easily surpass 100.000 euro per pile. Therefore employing bubble curtains efficiently whilst being sure of meeting the relevant regulations is important. This leads to the need for well validated models to predict the noise reduction of these bubble curtains.

Sound mitigation predictions can only be made if the hydrodynamical properties of the bubble curtain are known either from measurements or based on modelling. Hydrodyanimcal modelling of bubble plumes is therefore a crucial aspect and also the main focus of this paper. \citet{commander1989linear} showed that the sound attenuation depends on the void fraction and the bubble size distribution in the bubble curtain. The input to sound mitigation models should thus at least contain the local void fraction distribution and the local bubble size distribution. In order to optimize the use of bubble curtains these crucial properties should ideally be found by time efficient models. This has led to growing interest in combining hydrodynamical integral bubble plume models with sound models. \citet{bohne2019modeling} modelled the bubble curtain using an integral approach that provides a void fraction distribution while the bubble size distribution is assumed constant and taken from a measurement. On this basis, these authors derived the local sound speed based on \citet{commander1989linear} to finally obtain a prediction of the attenuation of the sound. In a later publication, \citet{bohne2020development} developed a model for the prediction of the bubble size distribution in a round plume and this distribution was then used as input to the straight plume model of \citet{bohne2019modeling}. \citet{peng2021study} incorporated the model of \citet{bohne2020development} into their sound propagation model for pile driving. 

The integral bubble plume models adopted in the recent sound mitigation models have been a topic of research for a long time. \citet{ditmars1975analysis} presented plume models for both point and line source plumes with a focus on water quality in lakes. \citet{mcdougall1978bubble} investigated the plumes arising from oil-well blow-outs. They derived two different round plume models which they solved using a power series. The work of \citet{milgram1983mean} provides a very complete description of a round plume model, including a clear derivation of initial conditions and a review of all the parameters involved. Moreover, they also verified their model using different experimental datasets. In relation to deep lake aeration \citet{wuest1992bubble} developed a model for round bubble plumes  which incorporated mass transport from the bubbles to the surrounding water. More recently, \citet{chu2017bubble,chu2019multiphase} used their round bubble plume model to estimate the intrusion and neutral height of a bubble plume in a stratified environment. As noted by \citet{dissanayake2021bubble}, the development of bubble plume models for straight plumes (originating from a line source) lacks behind as compared to round plumes. They themselves developed a planar plume model for use in stratified environments. \citet{fannelop1991surface} derived an analytical solution for planar plumes, although their research was more focused on the induced surface and recirculating flows rather than on the plume itself. Also the work of \citet{bohne2019modeling} deals with straight bubble plumes. Their model employs a virtual source in line with \citet{brevik1999role}, and uses the same parameters as \citet{cederwall1970analysis} to fit the model to the data. All of these straight plume models have in common that they assume an average flow field close to the nozzles including initial conditions that are averaged in the homogeneous direction. Nozzle geometry and the flow field close to the nozzle are however known to have an important impact on the bubble size distribution \citep[e.g.][]{ponasse1998bubble,voit1987calculation}, which can not be captured using this approach. Part of this paper focuses on combining straight plume modelling with capturing the important flow field characteristics close to the nozzle such that the bubble size distribution can be modelled throughout the entire domain. Further closing in on the gap between round and straight plume modelling, by now being able to continuously model the bubble size distribution in a straight plume.\\

Also the availability of reference data is limited for planar plumes. \citet{kobus1968analysis} reports measurements for both round and planar plumes, the latter conducted in a $1\,\mathrm{m}\times 10\,\mathrm{m}$ wide and $2\,\mathrm{m}$ deep tank at different air flowrates. They measured the water velocity, the width and the total velocity of the bubbles. In the BORA project (\citet{chmelnizkij2016schlussbericht}) planar bubble plumes have been investigated in a round tank with a diameter of $5\,\mathrm{m}$ and a depth of $4.8\,\mathrm{m}$. They measured the bubble size distributions and the void fraction distribution as well as the total rising velocity of the bubbles. Finally, in our previous work (\citet{beelen2023situ}), we provided data on the void fraction distribution, the growth rate and on the bubble size distributions in planar bubble plumes in a $4\,\mathrm{m}\times 220\,\mathrm{m}$ wide and $3.6\,\mathrm{m}$ deep basin. 

This paper has two major aims: To provide a dataset which can be used for verification of bubble plume models, and to present a model that predicts the relevant hydrodynamical properties and which is validated against our data. We start with an overview of the experimental setup in section \ref{sec:Experiments}. Then we derive the model in section \ref{sec:model}. The model is verified using the experimental data in section \ref{sec:results}, then we discuss the main model results and assumptions in section \ref{sec:Discussion}, followed by our conclusions in section \ref{sec:conclusions}.

\section{Experimental setup} \label{sec:Experiments}

\begin{figure}[!ht]
\centering
\includegraphics[width=\columnwidth]{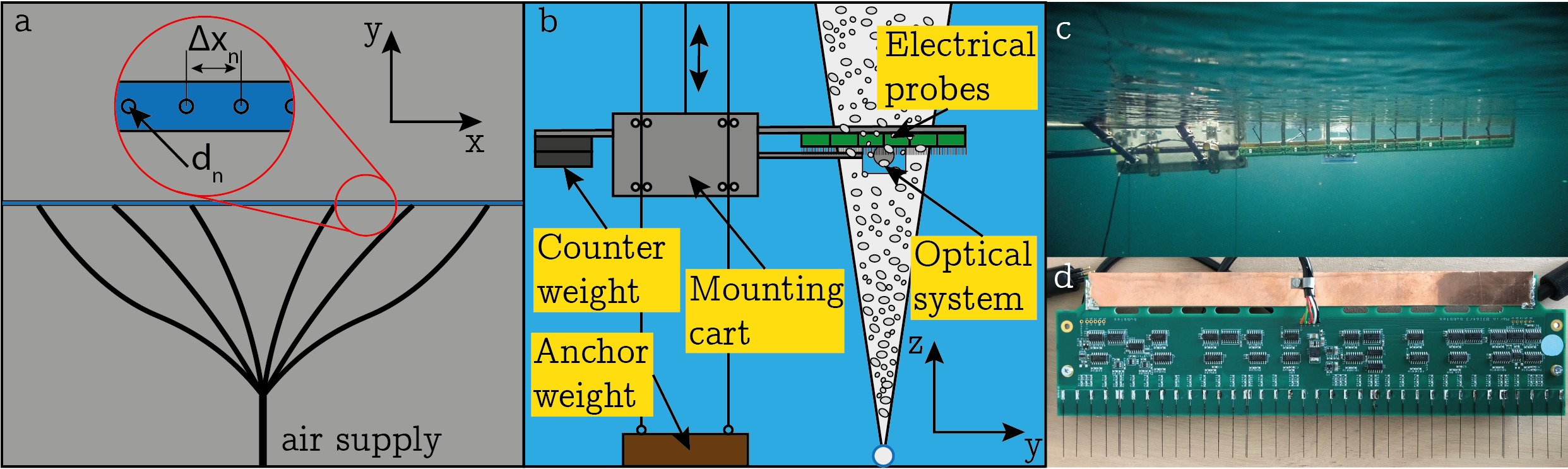}
\caption{a) Schematic overview of air supply and the hose in the OB b) Schematic overview of the mounting cart c) Image of the cart just below the waterline d) One of six PCBs used before mounting} \label{fig:Overview}
\end{figure}

The experimental results presented in this paper originate from two separate setups. Most of the data were obtained in the so-called Offshore Basin (OB) of the Dutch Maritime Research Institute (MARIN). The OB is a large freshwater tank with dimensions of $40\,\mathrm{m}\times 31\,\mathrm{m}$ wide and $10 \,\mathrm{m}$ deep. A PVC pipe with inner diameter $d_{\mathrm{pipe,in}}=20\,\mathrm{mm}$ and outer diameter $d_{\mathrm{pipe,out}}=25\,\mathrm{mm}$ was placed at the bottom of the tank at $z_0=-10\,\mathrm{m}$ covering the entire width of $31\,\mathrm{m}$. Drilled holes with diameter $d_{\mathrm{n}}=2\,\mathrm{mm}$ every $\Delta x_{\mathrm{n}}=100\,\mathrm{mm}$ along this pipe function as the air release nozzles. Every $5\,\mathrm{m}$ the pipe is supplied with air so that pressure drops within the pipe are minimized (see Figure \ref{fig:Overview}a). The air is supplied using a Berko diesel compressor, and the normal air flowrate is measured by a CS instruments VA 400 flow sensor.
The measurement systems (to be described in the next sections) are mounted on a cable controlled cart that is anchored to the bottom and can be moved to different heights (see Figure \ref{fig:Overview}b). Additional measurements carried out in MARIN's smaller Concept Basin (CB $4\,\mathrm{m}\times 220\,\mathrm{m}$ wide and $3.6\,\mathrm{m}$ deep) have been extensively described in \citet{beelen2023situ}.

\subsection{Electrical probes}\label{subsec22}
We employ a conductivity based technique to measure the void fraction distribution. Each individual probe consists of a needle that is $40\,\mathrm{mm}$ long with a diameter of $0.2\,\mathrm{mm}$. The body of the needle except for the exposed tip of approximately $1\,\mathrm{mm}$ is electrically insulated by a coating and a potential is applied. Whenever the tip is encapsulated in air the signal between the needle and the ground is broken, so that the bubbles can be detected. Several of these sensors along with their circuitry are assembled on a printable circuit board (PCB) and the use and calibration of these systems is described in detail in \citet{beelen2023situ}.

For the measurements in the OB, we used 6 PCBs carrying 40 equally spaced needles each to span a total width of $\Lambda=1.8\,\mathrm{m}$. The PCB's were operated in AC-mode to prevent tarnish from disrupting the measurements. Figure \ref{fig:Overview}c shows the electrical probes, as mounted during the measurements, just below the waterline and Figure \ref{fig:Overview}d shows a single PCB before mounting. 
The measurement campaign in the CB employed 2 of the same PCBs to span a total measurement distance  $\Lambda=0.6\,\mathrm{m}$. In this case, the PCBs were operated in DC mode. 
Although it is possible to infer the bubble size distribution based on the electrical probes, we solely used them for void fraction measurements in this paper. Instead, bubble size data reported here is based on the more accurate optical measurements described in the following.

\subsection{Optical system}\label{subsec23}

\begin{figure}[!ht]
\centering
\includegraphics[width=\columnwidth]{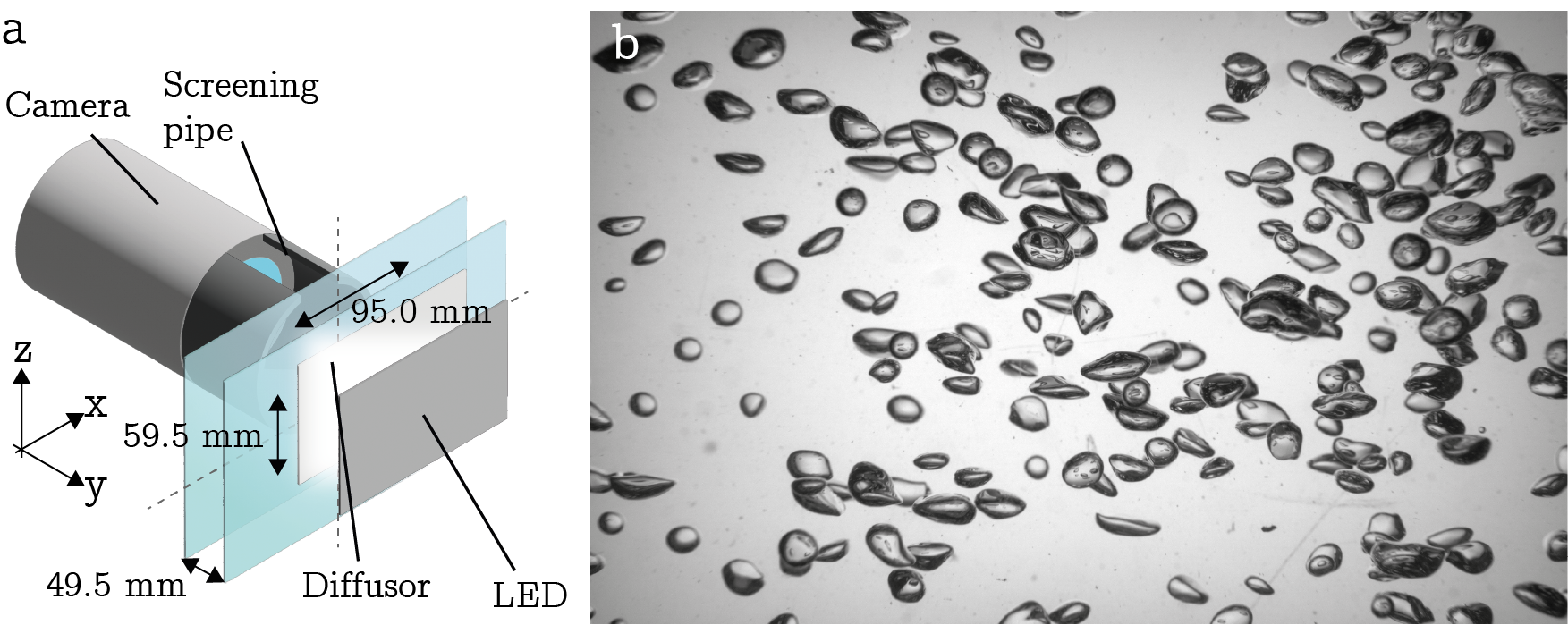}
\caption{a) Schematic overview of the optical system b) Typical image of bubbles in the bubble curtain} \label{fig:Optical}
\end{figure}

The optical system is schematically shown in Figure \ref{fig:Optical}a. The camera is placed inside a watertight casing. It registers the images of bubbles inside the measurement volume between two transparent plates. The volume is illuminated by a LED behind a diffusor plate and the space between the camera and the volume is bridged by a screening pipe to prevent out of focus bubbles from disturbing the images. In Figure \ref{fig:Optical}b a typical image as taken by the system is shown. The image algorithm used to retrieve the bubble size distribution along with more details on the optical system are described in \citet{beelen2023situ}. In the present study, the optical system has been used to determine the bubble size distribution.

\section{Continuous integral model for plumes originating from a linear nozzle array} \label{sec:model}

\begin{figure}[!ht]
\centering
\includegraphics[width=\columnwidth]{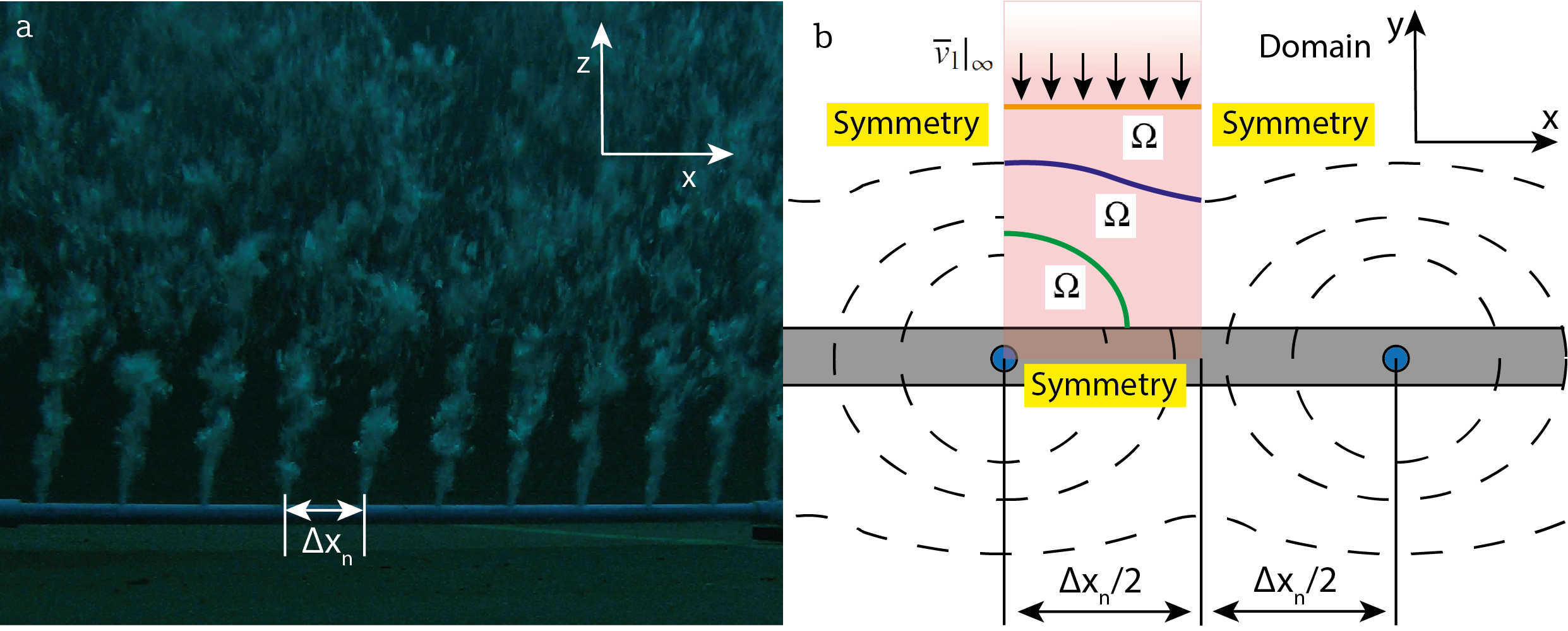}
\caption{a) Picture taken during the measurements of the air leaving the nozzles with the characteristic nozzle distance indicated b) Schematic overview of the domain, the dashed lines illustrate the geometry of the bubble curtain at different isocontours close to the nozzle. Within the domain a entrainment contour is given in green for a round plume, orange for a straight plume and blue for merging plumes} \label{fig:Domains}
\end{figure}

Planar bubble plumes typically originate from an array of round plumes emanating from individual nozzles along the pipe. These individual plumes then merge into a straight plume as can be seen in the snapshot in Figure \ref{fig:Domains}a. The image has been taken by an underwater drone during the measurements in the OB. The merging obviously depends on the spacing between the nozzles ($\Delta x_{\mathrm{n}}$), but also on other factors such as the nozzle diameter ($d_{\mathrm{n}}$) and the ambient gas flowrate per meter ($V^{*}_{\mathrm{ga}}$). Existing straight plume models do not take the initial round plume stage into account, but employ initial conditions representing averages along the pipe.  
The region close to the nozzle exit, however, is critical in determining the bubble size distribution throughout the entire bubble curtain and the relevant processes cannot be captured when using simple spatial averaging. A more faithful representation of the initial stage of the plume development can also be important for the local sound attenuation by the bubble curtain. This is because the average void fraction does generally not result in the average sound speed given the highly non-linear relation between the two. An added benefit of incorporating the initial stage is that we can make use of the well studied initial conditions for round plumes.

It is therefore our aim to provide a model that blends the initial round plume stage seamlessly into the development of the planar plume. Doing so requires a suitable domain based on which we can effectively describe the entire bubble curtain at every height. In Figure \ref{fig:Domains}b the proposed domain has been marked in red. Schematic contour lines representing typical void fraction or velocity distributions at different distances close to the nozzle illustrate the merging of the plumes. Taking advantage of the constant nozzle spacing allows us to employ symmetry along the pipe direction to incorporate all stages of the bubble curtain with limited computational effort. This enables us to build upon the model of \citet{bohne2020development} and to apply it continuously for a straight plume, overcoming the difficulties of the initial description of a straight plume. \citet{bohne2020development} used the standard integral plume assumptions, i.e. symmetry across the center line, steady mean fields, Gaussian distributions for the velocity and void fraction fields, and the Morton-Taylor-Turner entrainment hypothesis (\citet{morton1956turbulent}) with a constant entrainment factor. Additionally, integrated transport equations for the volume fractions and the average bubble volumes as derived by \citet{lehr2002bubble} and \citet{mewes2003two} are solved. Our proposed model follows this approach. In presenting the relevant equations, we will use the following coordinate system (see also Figure \ref{fig:Domains}a and b): In the horizontal plane, the $x$-axis runs along the pipe and the $y$-direction is perpendicular to it. The vertical ($z$-direction) is pointing upward with the origin $z = 0 $ located at the water surface. The representative domain describing the entire plume is defined within the symmetry lines which are located at the nozzle ($x=0$) and in the middle between two nozzles ($x=\frac{1}{2}\Delta x_{\mathrm{n}}$) and in line with the pipe ($y=0$). The domain thus spans from $0<x<\frac{1}{2}\Delta x_{\mathrm{n}}$ and $0<y<\infty$.

Using this domain instead of resorting to the conventional approach of considering dependencies on the radial distance or the spanwise coordinate for round and planar plumes, respectively, allows us to incorporate a smooth transition between the two limiting cases. Since we consider a steady state for the mean field, we will start our derivations from the Reynolds averaged equations.

\subsection{Velocity and void fraction field}
\label{subsec:3.4}

\begin{figure}[!ht]
\centering
\includegraphics[width=0.75\textwidth]{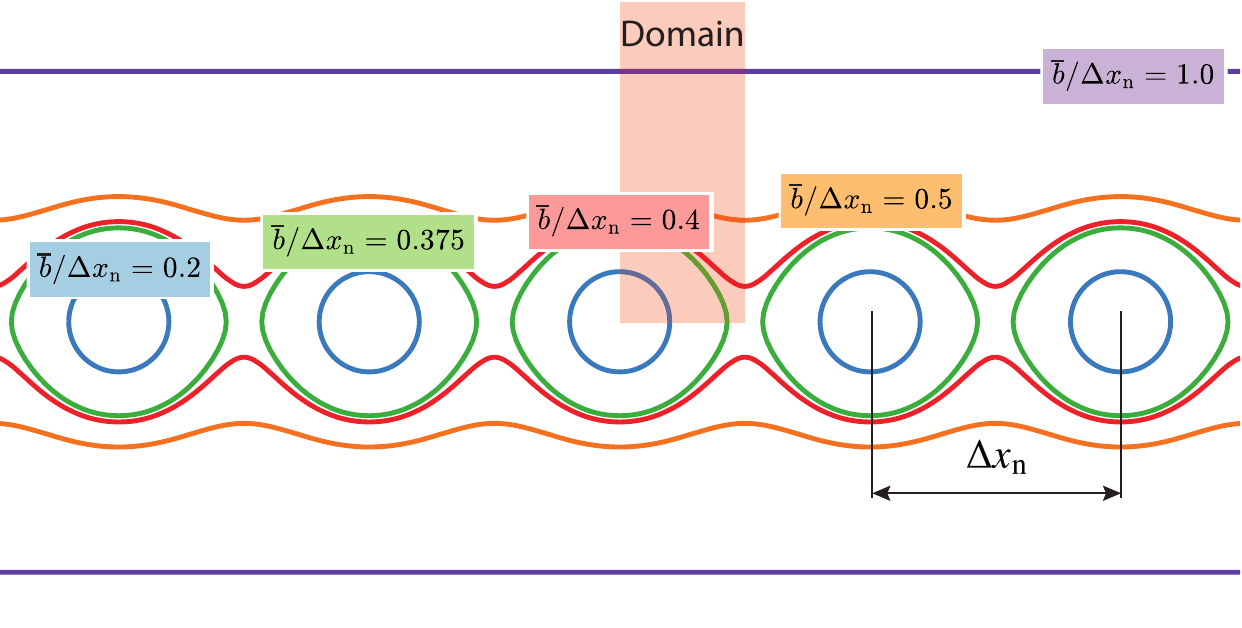}
\caption{Isocontours of the velocity field defined in Eq. \ref{eq:vel2} at $\overline{w}_{\mathrm{l}} = 1/e \overline{w}_{\mathrm{lm}}$ and for different values of $\overline{b}/\Delta x_{\mathrm{n}}$} \label{fig:Velcontour}
\end{figure}

The bubble curtain originates from nozzles in a hose as individual round plumes, and then transitions to a straight plume and distributions are commonly approximated as Gaussian with radial and linear symmetry respectively. A simple normalized summation of round plumes is consistent with both of these limiting regimes and complies with the symmetry conditions described above. We therefore use the following expression for the time-averaged vertical velocity field $\overline{w}_{\mathrm{l}}$:

\begin{equation}
\begin{split}
\overline{w}_{\mathrm{l}}=\overline{w}_{\mathrm{lm}}f_{\mathrm{w}}\\
f_{\mathrm{w}}= \frac{1}{\max(f'_w)}f'_w \quad \textrm{with } f'_w = \sum_{j=-\infty}^{\infty}{\exp\left(-((x-j\Delta x_{\mathrm{n}})^2 +y^2)/\overline{b}^2\right)}, 
\label{eq:vel2}
\end{split}
\end{equation}
where $\overline{b}$ is the Gaussian standard deviation and $\overline{w}_{\mathrm{lm}}$ is the velocity above a nozzle (i.e. the velocity at $x=j\Delta x_{\mathrm{n}}$, with $j\in\mathbb{Z}$ and $y=0$). From now on we will call $\overline{w}_{\mathrm{lm}}$ the centerpoint velocity. As shown in Figure \ref{fig:Velcontour}, Eq. \ref{eq:vel2} describes an array of round Gaussian plumes if $\Delta x_{\mathrm{n}}\gg \overline{b}$, while the limit $\Delta x_{\mathrm{n}}\ll \overline{b}$ recovers a straight plume with a Gaussian distribution with the smooth transition between these two cases at intermediate $\Delta x_{\mathrm{n}}$. Note that the contour lines in Figure \ref{fig:Velcontour} correspond to $\overline{w}_{\mathrm{l}} = 1/e \overline{w}_{\mathrm{lm}}$. For higher contour levels, the shape transition appears at higher values of $\overline{b}/\Delta x_{\mathrm{n}}$. 

Analogously, a time averaged void fraction field can be defined according to

\begin{equation}
\begin{split}
\overline{\epsilon}_{\mathrm{gk}}=\overline{\epsilon}_{\mathrm{gmk}}\frac{1}{\text{max}(f_{\epsilon})}f_{\epsilon}\\
f_{\epsilon}=\sum_{n=-\infty}^{\infty}{\exp\left(-((x-n\Delta x_{\mathrm{n}})^2 +y^2)/\overline{b}^2\lambda^2 \right)}.
\label{eq:void2}
\end{split}
\end{equation}
Here, the parameter $\lambda$ is the ratio between the standard deviations of the void fraction and velocity fields and accounts for the fact that the void fraction distribution is generally somewhat narrower.
The subscript $k$ is used to refer to the distributions of "small" bubbles (1) or "large" bubbles (2) (see section \ref{subsec:voidfrac}), and $\epsilon_{\mathrm{gmk}}$ is called the centerpoint void fraction.

\subsection{Momentum conservation}
We start from the Reynolds-averaged momentum conservation equation for multiphase flow as given for example by \citet{wang2005numerical}. Since the gas density $\rho_g$ is much smaller than that of the liquid ($\rho_l$), the momentum is almost exclusively carried by the water, such that it is sufficient to consider the momentum equation for the liquid only. We further note that for typical void fractions of the order of a few percent, fluctuations in the liquid fraction $\epsilon_{\mathrm{l}}$ are negligibly small. In a steady state the momentum conservation in the $z$-direction then reads

\begin{equation}
    \begin{split}
    \pder{}{x}\left(\overline{\epsilon}_{\mathrm{l}}\rho_{\mathrm{l}} \overline{u}_{\mathrm{l}}\overline{w}_{\mathrm{l}}+\overline{\epsilon}_{\mathrm{l}}\rho_{\mathrm{l}} \overline{u'_{\mathrm{l}}w'_{\mathrm{l}}}\right) + \pder{}{y}\left(\overline{\epsilon}_{\mathrm{l}}\rho_{\mathrm{l}} \overline{v}_{\mathrm{l}}\overline{w}_{\mathrm{l}}+\overline{\epsilon}_{\mathrm{l}}\rho_{\mathrm{l}}\overline{v'_{\mathrm{l}}w'_{\mathrm{l}}}\right)+ \pder{}{z}\left(\overline{\epsilon}_{\mathrm{l}}\rho_{\mathrm{l}} \overline{w}_{\mathrm{l}}\overline{w}_{\mathrm{l}}+\overline{\epsilon}_{\mathrm{l}}\rho_{\mathrm{l}}\overline{w'_{\mathrm{l}}w'_{\mathrm{l}}}\right)=-\pder{\overline{p}}{z}-\overline{\epsilon}_{\mathrm{l}}\rho_{\mathrm{l}} g.
    \end{split}
    \label{eq:ReynodsavgMOM}
\end{equation}
Here, the liquid velocity in the $x$, $y$ and $z$-direction is $u_{\mathrm{l}}$,  $v_{\mathrm{l}}$ and $w_{\mathrm{l}}$ respectively. We use the overbar to denote temporal averaging and the prime indicates the fluctuating part, such that e.g. $u = \overline{u}+u'$. In the absence of ambient stratification, the mean pressure gradient is given by the hydrostatic pressure distribution
$\pder{\overline{p}}{z}=-\rho_{\mathrm{l}}g$, with $g$ denoting the gravitational acceleration. With the gas void fraction $\overline{\epsilon}_{\mathrm{g}} = 1- \overline{\epsilon}_{\mathrm{l}}$ the right hand side of Eq. \ref{eq:ReynodsavgMOM} reduces to $\overline{\epsilon}_{\mathrm{g}}\rho_l g$. When integrating Eq. \ref{eq:ReynodsavgMOM} over the domain indicated in Figure 3b the transport related to the mean field in the first two terms on the left-hand side vanishes since $\overline{u}_{\mathrm{l}}(0,y)=\overline{u}_{\mathrm{l}}(\Delta x_{\mathrm{n}}/2,y)=0$ and $\overline{v}_{\mathrm{l}}(x,0)=0$ due to symmetry and and $\overline{w}_{\mathrm{l}}(x,\infty)\rightarrow 0$. Additionally, in the absence of a mean gradient across the symmetry boundary, also the contribution of the fluctuating terms is zero for these two terms. We assume that the remaining fluctuating term is proportional to the mean field according to $\overline{w'w'}= (\gamma-1)\overline{w}\overline{w}$, where $\gamma$ is the momentum amplification factor, such that integration of Eq. \ref{eq:ReynodsavgMOM} yields

\begin{equation}
    \pder{}{z}\int_0^\infty{\int_0^{\Delta x_{n}/2}{\left(\gamma\overline{\epsilon}_{\mathrm{l}}\rho_{\mathrm{l}} \overline{w}_{l}\overline{w}_{l}\right)}}dxdy=\int_0^\infty{\int_0^{\Delta x_{n}/2}{\overline{\epsilon}_g\rho_{\mathrm{l}} g}}dxdy.
\label{eq:Momentumresult}
\end{equation}

As outlined in \citet{milgram1983mean}, the amplification factor for bubble plumes approaches that of solutal plumes in case the vertical extent is large compared to a typical bubble spacing. With typical bubble spacings of several centimeters and heights of meters, this applies here. In this case, it is common to use the approximation $\gamma = 1$, which we therefore also adopt here consistent with \citet{bohne2019modeling} and \citet{cederwall1970analysis}. However, the results are not very sensitive to this choice and setting $\gamma = 1.1$ does not make a significant difference.
 Defining the integral liquid momentum flux $M_{\mathrm{l}}=\int_0^\infty{\int_0^{\Delta x_{n}/2}{\left(\gamma\overline{\epsilon}_{\mathrm{l}}\rho_{\mathrm{l}} \overline{w}_{l}\overline{w}_{l}\right)}}dxdy$ and the integral buoyancy $B=\int_0^\infty{\int_0^{\Delta x_{n}/2}{\overline{\epsilon}_g\rho_{\mathrm{l}} g}}dxdy$ reduces the integral form of Eq. \ref{eq:Momentumresult} to

\begin{equation}
    \pder{M_{\mathrm{l}}}{z}=B.
\label{eq:Momentumresultsimp}
\end{equation}

\subsection{Mass conservation}

The steady-state time averaged mass conservation equation for the liquid phase is given by (see e.g. \citet{wang2005numerical})

\begin{equation}
    \pder{}{x}(\overline{\epsilon}_{\mathrm{l}}\rho_{\mathrm{l}} \overline{u}_{\mathrm{l}})+\pder{}{y} (\overline{\epsilon}_{\mathrm{l}}\rho_{\mathrm{l}} \overline{v}_{\mathrm{l}})+\pder{}{z}(\overline{\epsilon}_{\mathrm{l}}\rho_{\mathrm{l}} \overline{w}_{\mathrm{l}})=0.
    \label{eq:massconserv1}
\end{equation}

When integrating Eq. \ref{eq:massconserv1} over the symmetry domain contributions from all symmetry boundaries are zero, leading to 

\begin{equation}
    \frac{1}{2}\Delta x_{\mathrm{n}}\rho_{\mathrm{l}}\left.\overline{v}_{\mathrm{l}}\right\vert_{\infty}
    +\int_0^\infty{\int_0^{\Delta x_{\mathrm{n}}/2}{\pder{}{z}(\overline{\epsilon}_{\mathrm{l}}\rho_{\mathrm{l}} \overline{w}_{\mathrm{l}})}}dxdy=0,
    \label{eq:mass_integrals}
\end{equation}
where $\left.\overline{v}_{\mathrm{l}}\right\vert_{\infty}$ is the unknown entrainment velocity at $y \rightarrow \infty$. We can express the entrained flux by the flux across any contour $\Omega$ inside the domain (in Figure \ref{fig:Domains}b some example contours can be seen for different plume shapes), according to
\begin{equation}
     \frac{1}{2}\Delta x_{\mathrm{n}}\rho_{\mathrm{l}}\left.\overline{v}_{\mathrm{l}}\right\vert_{\infty}=\rho_{\mathrm{l}}\int_{\Omega}\overline{\mathbf{u}}_{\mathrm{h}}\cdot \mathbf{n} d\Omega, 
\end{equation}
which holds as long as the vertical velocity at the location of $\Omega$ is negligibly small. Here, $\overline{\mathbf{u}}_{\mathrm{h}}$ is a vector containing the horizontal velocity components, $\mathbf{n}$ is the outward pointing normal vector of the contour element $d\Omega$. We can then apply the entrainment hypothesis according to

\begin{equation}
    \rho_{\mathrm{l}}\int_{\Omega}\overline{\mathbf{u}}_{\mathrm{h}}\cdot \mathbf{n} d\Omega = \alpha \rho_{\mathrm{l}} \overline{w}_{\mathrm{lm}}P, 
    \label{eq:Entrainflux}
\end{equation}
where $P$ is the length of an appropriately chosen isocontour of $\overline{w}_{\mathrm{l}}$.
The choice of $P$ needs to be consistent with the definition of the entrainment coefficient $\alpha$, that relates the entrainment velocity to the vertical velocity in the center of the plume $\overline{w}_{\mathrm{lm}}$. We adopt a definition of $\alpha$ based on the spreading of the Gaussian standard deviation $\overline{b}$, such that the corresponding isocontour $P$ is given by $\overline{w}_{\mathrm{l}}=\overline{w}_{\mathrm{l}}(0,\overline{b}) = 1/e \overline{w}_{lm}$. Eq. \ref{eq:Entrainflux} smoothly interpolates between the limiting cases of a round plume, for which $P$ is the length of a quarter circle, to a planar plume where $P = \Delta x_{\mathrm{n}}/2$. Introducing the integral liquid mass flux $Q_{\mathrm{l}}=\int_0^\infty{\int_0^{\Delta x_{\mathrm{n}}/2}{(\overline{\epsilon}_{\mathrm{l}}\rho_{\mathrm{l}} \overline{w}_{\mathrm{l}})}}dxdy$, the integrated form of Eq. \ref{eq:massconserv1} is then given by
\begin{equation}
    \pder{Q_{\mathrm{l}}}{z}=\alpha \overline{w}_{\mathrm{lm}}\rho_{\mathrm{l}}P.
    \label{eq:massresult}
\end{equation}

\paragraph{Entrainment factor}\label{subsec:entrain}

\begin{figure}[!ht]
 \centering
 \includegraphics[width=\columnwidth]{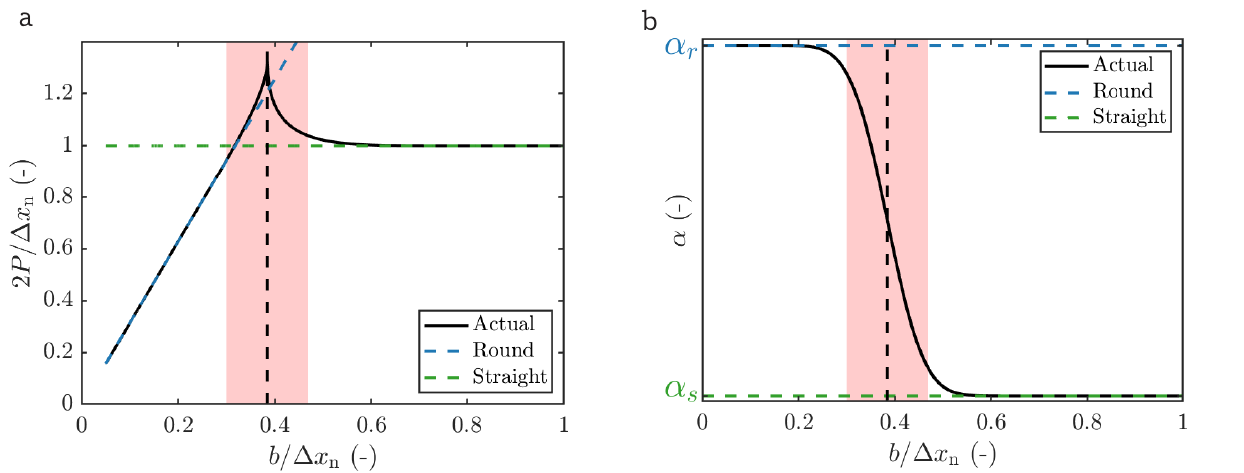}
 \caption{a) The normalized perimeter of the bubble curtain in the symmetry domain as a function of the normalized standard deviation. b) The resulting entrainment factor by applying Eq. \ref{eq:Entrainequation}, in this example $\alpha_{\mathrm{r}}=0.1$ and $\alpha_{\mathrm{s}}=0.094$} \label{fig:entrainexample2}
 \end{figure}

The spread of the bubble curtain is determined by the entrainment factor. The bubble plume originates from round plumes, for which the entrainment factor $\alpha_r=0.1$ differs from the value applicable for straight plumes ($\alpha_s$, see section \ref{subsec:entrainres} for details). We utilize the perimeter to distinguish between these two regimes. Initially, the perimeter grows as the perimeter of a quarter circle $P=\frac{1}{2}\pi \overline{b}$, up to the point at which individual plumes merge, which leads to a faster increase in $P$ as shown in Figure \ref{fig:entrainexample2}a. Finally when the bubble curtain is straight the perimeter reduces to $P=\frac{\Delta x_{\mathrm{n}}}{2}$. Based on  Figure \ref{fig:entrainexample2}a, the transition from round to planar is centered around $\overline{b}_{\mathrm{trans}}/\Delta x_{\mathrm{n}}\approx 0.385$ and we therefore smoothly transition the effective entrainment coefficient according to 
\begin{equation}
    \alpha(\overline{b}/\Delta x) =\frac{1}{2}\left(\alpha_{\mathrm{s}}-\alpha_{\mathrm{r}}\right)\left[\text{erf}\left(2\frac{\overline{b}/\Delta x_{\mathrm{n}}-0.385}{0.17}\right)+1\right]+\alpha_{\mathrm{r}},
\label{eq:Entrainequation}
\end{equation}
where $\overline{b}_{\mathrm{reg}}/\Delta x_{\mathrm{n}}\approx 0.17$ is the normalized width of the transition region (shaded red in Figure \ref{fig:entrainexample2}). Eq. \ref{eq:Entrainequation} is plotted vs. $\overline{b}/\Delta x_{\mathrm{n}}$ in Figure \ref{fig:entrainexample2}b.

\subsection{Void fraction and bubble volume}
\label{subsec:voidfrac}
The treatment of the bubble phase starts from the assumption that the total bubble population can be described by a superposition of a `small' -bubble  and a `large'-bubble distribution \citep{lehr2002bubble}, which will be referred to by subscripts "1" and "2", respectively. We further consider variations of the bubble size distribution with respect to the vertical coordinate only. 
 
The number density $n_1$ of the small bubbles that contribute to the small bubble void fraction $\overline{\epsilon}_{\mathrm{g1}}$, is assumed to be log-normally distributed according to 

\begin{equation}
    n_1(\nu)=\frac{\overline{\epsilon}_{\mathrm{g1}}}{\tilde{\nu}_1}\frac{1}{\nu\xi\sqrt{2\pi}}\exp{\left[-\frac{1}{2\xi^2}\ln{\left(\exp{(\xi^2/2)}\frac{\nu}{\tilde{\nu}_1}\right)^2}\right]},
    \label{eq:numbdistri1}
\end{equation}
where $\nu=\frac{4}{3}\pi a^3$ is the bubble volume and $a$ is the bubble radius. The mean bubble volume is denoted by $\tilde{\nu}_1$, where the tilde indicates the average over the bubble size distribution from here on. Following \citet{lehr2002bubble} we use $\xi=\frac{3}{2}$. Log-normal distributions are often observed to be the best fit in bubbly flows where the bubble size is determined by break-up and coalescence in turbulence \citep[e.g.][]{adetunji2017estimation,colella1999study,leibson1956rate,mandal2005comparative,senouci2018hydrodynamics}. \citet{lehr2002bubble} remarks that equal sized breakage is preferred for smaller bubbles as the interfacial forces are high and equal breakage minimizes the required inertial force of an eddy. Larger bubbles, on the other hand, prefer unequal breakage, since they tend to interact with relatively small eddies. The aforementioned mechanism "competes" with (1) small bubbles remaining stable as the inertial force exerted by an eddy is insufficient to break the surface tension force and (2) bubble coalescence. While these arguments do not determine the  exact shape of the distribution, they do explain the existence of a peak at a certain small-bubble size and the skewness due to remaining large bubbles in the distribution.

The number density $n_2$ of the larger bubbles is approximated by the exponential distribution

\begin{equation}
    n_2(\nu)=\frac{\overline{\epsilon}_{\mathrm{g2}}}{\tilde{\nu}_2}\frac{1}{\tilde{\nu}_2}\exp{-\frac{\nu}{\tilde{\nu}_2}},
\end{equation}
with mean bubble volume $\tilde{\nu}_2$ and void fraction $\overline{\epsilon}_{\mathrm{g2}}$. Close to the nozzle large bubbles are present, as is shown by the multitude of studies into the creation of primary bubbles at the orifice exit (\citep[e.g.][]{badam2007experimental, gaddis1986bubble, jamialahmadi2001study,kulkarni2005bubble,ramakrishnan1969studies}). \citet{lehr2002bubble} observed the presence of these larger bubbles in regions of high void fraction, and described their size distribution using an exponential distribution. 
Even though this approach is mostly empirical, the effectiveness of the assumption made by \citet{lehr2002bubble} is evident by the wide spread use and successful application of their model for bubble columns. In practice, the exponential distribution is only relevant close to the nozzle due to the swift break-up of the larger bubbles, as will be discussed in section \ref{subsec:5.1}.

The total bubble number density of all bubbles is $n=n_1+n_2$, and the probability density function is given by

\begin{equation}
    f_{\mathrm{num}}=\frac{n}{\int_0^{\infty}{nda}}, 
    \label{eq:numbdistri}
\end{equation}
In order to characterize the distribution of the total bubble volume across different bubble sizes, it is useful to weigh $f_{\mathrm{num}}$ by the bubble volume according to 
\begin{equation}
    f_{\mathrm{v}}=\frac{f_{\mathrm{num}}\frac{4}{3}\pi a^3}{\int_0^\infty {f_{\mathrm{num}} \frac{4}{3}\pi a^3 da}}
\end{equation}
The two bubble size distributions "1" and "2" interact with each other through break up and coalescence. Adopting the expressions provided in \citet{lehr2002bubble}, the break up kernels for small ($Z_1$) and large bubbles ($Z_2$) are given in Eqs. \ref{eq:smallbreakup} and \ref{eq:largebreakup}, respectively.

\begin{equation}
    Z_1(\tilde{\nu}_1)=0.6082\left(\frac{\rho_{\mathrm{l}}}{\sigma}\right)\overline{\varepsilon}^{19/15}\tilde{\nu}_1^{5/9} \exp\left(\frac{-0.8565}{\tilde{\nu}_1^{0.5}}\left(\frac{\sigma}{\rho_{\mathrm{l}}}\right)^{9/10} \frac{1}{\overline{\varepsilon}^{3/5}}\right)
    \label{eq:smallbreakup}
\end{equation}

\begin{equation}
    Z_2(\tilde{\nu}_2)=0.2545\left(\frac{\sigma}{\rho_{\mathrm{l}}}\right)^{2/5} \overline{\varepsilon}^{1/15}\frac{1}{\tilde{\nu}_2^{4/9}}
    \label{eq:largebreakup}
\end{equation}
Here, $\sigma$ denotes the surface tension at the water air interface. Bubbles break up due to interaction with turbulent eddies on the order of the bubble size with sufficient energy (\citet{hinze1975turbulence}). A measure for the energy in these turbulent eddies is provided by the local temporal average energy dissipation rate $\overline{\varepsilon}$, which therefore needs to be modelled. In the most general case, $\overline{\varepsilon}$ depends on all spatial coordinates, but this complexity cannot be represented in our model, which is essentially one dimensional. Following \citet{bohne2020development} we therefore simplify the distribution of $\overline{\varepsilon}$ to

\begin{equation}
        \overline{\varepsilon}(x,y,z) = \left\{
    \begin{array}{ll}
        \langle\langle{\overline{\varepsilon}}\rangle\rangle(z) & \Gamma(x,y,z) > 50\%  \\
        0 & \Gamma(x,y,z) \leq 50\% ,
    \end{array}
\right.
\label{eq:Dissipationratedistri}
\end{equation}
where $\langle\langle{\overline{\varepsilon}}\rangle\rangle$ is the horizontally averaged (denoted by $\langle\langle\cdot\rangle\rangle$) dissipation rate and hence a function of $z$ only, and its value is zero in regions where the intermittency factor $\Gamma$ (percentage of time the flow is turbulent) is smaller than $50\%$. For round plumes an intermittency factor of $50\%$ is reached at $\overline{b}_{\Gamma = 50\%} =\sqrt{2}\overline{b} $ \citep{lee2003turbulent}. To generalise this, we use us the corresponding velocity isocontour and assume that $\Gamma(x,y,z) > 50\%$ where $\overline{w}_{\mathrm{l}}>\overline{w}_{\mathrm{l}}(0,\sqrt{2}\overline{b},z)$. At the value of $\lambda = 0.85$ employed here, 98\% of all bubbles are contained within these bounds, which further supports this choice. Furthermore, the break-up rate at low levels of $\varepsilon$ far from the centerline is small.

For the mean dissipation rate we adopt the expression derived by \citet{zhao2014evolution}. Their model includes a correction term for the region close to the nozzle and reads

\begin{equation}
\langle\langle\overline{\varepsilon}\rangle\rangle=\text{min}\left(0.003\frac{\langle\langle\overline{w}_0\rangle\rangle^3}{d_{\mathrm{n}}}; c_{\varepsilon} \frac{\overline{w}_{\mathrm{lm}}^3}{\overline{b}}\right), 
\label{eq:meandissZhao}
\end{equation}
where $\langle\langle\overline{w}_0\rangle\rangle$ is the mean outflow velocity over the nozzle opening and $c_{\varepsilon}$ is a constant for which \citet{zhao2014evolution} specify a range between $0.024<c_{\varepsilon}<0.043$.

The rate of coalescence is largely determined by the overall centerpoint void fraction $\overline{\epsilon}_{\mathrm{gm}}=\overline{\epsilon}_{\mathrm{gm1}}+\overline{\epsilon}_{\mathrm{gm2}}$ as is reflected in the coalescence kernel $r_2$ between two bubbles with volumes $\nu^{\blacktriangle}$ and $\nu^{\blacktriangledown}$

\begin{equation}
    r_2(\nu^{\blacktriangle},\nu^{\blacktriangledown})=\frac{\pi}{4}\left(\left(\frac{6}{\pi}\nu^{\blacktriangle}\right)^{1/3}+\left(\frac{6}{\pi}\nu^{\blacktriangledown}\right)^{1/3}\right)u_{\mathrm{crit}}\exp\left(-\left(\frac{\epsilon_{g\mathrm{max}}^{1/3}}{\overline{\epsilon}_{\mathrm{gm}}^{1/3}}-1\right)^2\right).
\end{equation}
Here, we use $\epsilon_{g\mathrm{max}} = 0.6$ based on the maximum packing density of spheres (see \citet{millies1999interfacial}) and  $u_{\mathrm{crit}} = 0.08\,\mathrm{ms^{-1}}$ for the critical coalescence velocity, the same values have also been adopted in \citet{bohne2020development}. 

\paragraph{Slip velocity}
An important difference between bubble plumes and solutal plumes is the presence of a relative velocity of the bubbles with respect to the fluid. This manifests in the vertical transport where the total vertical velocity of the gas bubbles is given by $\overline{w}_{\mathrm{g,k}}=\overline{w}_{\mathrm{l}}+\overline{w}_{\mathrm{rel,k}}$. Note that $\overline{w}_{\mathrm{rel,k}}$ accounts for the slip velocity of the bubbles, but also potential other effects such as preferential sampling of flow regions and collective effects. Neither of these has a net effect in the horizontal directions, such that  $\overline{u}_{\mathrm{g,k}}=\overline{u}_{\mathrm{l}}$ and $\overline{v}_{\mathrm{g,k}}=\overline{v}_{\mathrm{l}}$. We will neglect a potential dependence of $\overline{w}_{\mathrm{rel,k}}$ on the position within the bubble curtain \citet{deen2000comparison} and also assume it to be independent of bubble size. This latter  choice is in line with \citet{bohne2020development} and the fact that for single bubbles the terminal velocity, in the relevant size range of $1 <a <10\,\mathrm{mm}$, is approximately constant, see e.g. \citet{wuest1992bubble}. Based on the fit with our data, we adopted a value of

\begin{equation}
    \overline{w}_{\mathrm{rel,k}}= \overline{w}_{\mathrm{rel}}=0.8\,\mathrm{ms^{-1}},
\end{equation}
whereas \citet{bohne2020development} used $\overline{w}_{\mathrm{rel,k}}=0.3\,\mathrm{ms^{-1}}$. The large difference between these values will be evaluated extensively in paragraph  \ref{subsec:slipvel}.

\paragraph{Gas-liquid mass transfer}
In principle, the void fraction and bubble volume can also change due to gas exchange between the bubbles and the surrounding liquid. Even though bubble plumes are for example used for the aeration of water masses \citep{singleton2006designing,wuest1992bubble}, we do not expect an important role for gas exchange for the applications in mind. To substantiate this, we employ a simple model for the gas exchange. We assume the water to be well mixed such that the dissolved air concentration is constant and at a 100\% saturation level at the water surface ($c_{\mathrm{air}}=c_{\mathrm{s,air}}(p=1\,\mathrm{atm})$). 
Note that considering air as a mixture instead of its components individually implicitly assumes that the change in the composition of the gas inside the bubble is negligible.
The local gas flux through a bubble surface is than given by $k_{\mathrm{c,air}}(c_{\mathrm{s,air}}-c_{\mathrm{air}})$. Here, $k_{\mathrm{c,air}}$ is a mass transfer coefficient and the local saturation concentration ($c_{\mathrm{s,air}}$) depends on the pressure inside the bubble, which for sufficiently large bubbles can be assumed to be equal to the surrounding pressure. The molar mass flow of gas from the bubble is then given by $\mathcal{M}_{\mathrm{m}}A_{\mathrm{bub}}k_{\mathrm{c,air}}(c_{\mathrm{s,air}}-c_{\mathrm{air}})$, with $\mathcal{M}_{\mathrm{m}}$ the molar mass of air and $A_{\mathrm{bub}}$ the surface area of a bubble. Assuming a constant rise velocity of $\overline{w}_{\mathrm{g}}$, we find a change in bubble volume over the height
\begin{equation}
\frac{dV_{\mathrm{bub}}}{dz}=-\frac{\mathcal{M}_{\mathrm{m}}A_{\mathrm{bub}}}{\rho_{\mathrm{g}}\overline{w}_{\mathrm{g}}}k_{\mathrm{c,air}}(c_{\mathrm{s,air}}-c_{\mathrm{air}}).   
\label{eq:Massflux}
\end{equation}
Integrating this equation numerically for a bubble with $d_{\mathrm{eq}}=5\,\mathrm{mm}$, $\overline{w}_{\mathrm{g}}=1\,\mathrm{ms^{-1}}$ and using the values proposed by \citet{wuest1992bubble} for the mass transfer coefficient and the saturation concentration (at a temperature of $20^{\circ}C$) we find a change in bubble volume of less than 3\%, or a change in the bubble diameter of approximately 1\% in the OB. 
These estimates are likely to overestimate the true effect, since the dissolved gas is dragged along in the plume leading to oversaturation and therefore bubble growth at lower depths. The effect of mass transport is therefore indeed negligible in the present configuration. If needed though, the present model can easily be expanded to account for mass transfer, by including equation \ref{eq:Massflux} along with a conservation equation for the dissolved gas concentration as outlined in \citet{wuest1992bubble}.

\subsubsection{Transport equations for the void fraction}
The transport of the small bubble void fraction is given by (\citet{bohne2020development} and \citet{lehr2002bubble}). The transport equation for the small bubble void fraction follows from integrating the population balance equations over all bubble volumes. In doing so, it is assumed that the distribution of the so-called daughter bubbles after break-up equals that of the small bubbles (Eq. \ref{eq:numbdistri1}). Coalescence is taken into account through the coalescence kernel function which has been investigated experimentally by \citet{lehr2002bubble}. The resulting transport equation for the void fraction $\overline{\epsilon}_{\mathrm{g1}}$ carried by the small bubbles is then given by 

\begin{equation}
\begin{split}
\pder{}{x}\left(\rho_{\mathrm{g}}\overline{\epsilon}_{\mathrm{g1}}\overline{u}_{\mathrm{l}}\right)+\pder{}{y}\left(\rho_{\mathrm{g}}\overline{\epsilon}_{\mathrm{g1}}\overline{v}_{\mathrm{l}}\right)+\pder{}{z}\left(\rho_{\mathrm{g}}\overline{\epsilon}_{\mathrm{g1}}\overline{w}_{\mathrm{g1}}\right)=\\Z_2\rho_g\overline{\epsilon}_{g2}-0.9024r_2(\tilde{\nu}_2,5\tilde{\nu}_1)\rho_g\frac{\overline{\epsilon}_{g2}\overline{\epsilon}_{g1}}{\tilde{\nu}_2}-3.1043r_2(\tilde{\nu}_1,\tilde{\nu}_1)\rho_g\frac{\overline{\epsilon}_{\mathrm{g1}}^2}{\tilde{\nu}_2}.
\end{split}
\label{eq:smallbubsgoverning}
\end{equation}
Here, the individual terms on the right-hand side respectively relate to an increase of the small bubble void fraction due to the break-up of the larger bubbles into smaller bubbles ($Z_2$), a decrease due to coalescence of small bubbles with larger bubbles ($r_2(\tilde{\nu}_2,5\tilde{\nu}_1)$) and an additional decrease due to the coalescence of small bubbles with other small bubbles to form a large bubble ($r_2(\tilde{\nu}_1,\tilde{\nu}_1)$). The void fraction does not depend on the break up of small bubbles since in this case the resulting bubbles remain in the same category.

Integrating both sides of Eq. \ref{eq:smallbubsgoverning}, and using the same symmetry arguments as for the momentum and mass conservation gives
 
\begin{equation}
\begin{split}
\pder{}{z}\int_{0}^{\infty}{\int_{0}^{\Delta x_{\mathrm{n}}/2}{\rho_g\overline{\epsilon}_{g1}(\overline{w}_{\mathrm{l}}+\overline{w}_{\mathrm{rel},1})dx}dy}=\int_{0}^{\infty}{\int_{0}^{\Delta x_{\mathrm{n}}/2}{Z_2\rho_g\overline{\epsilon}_{g2}dx}dy}\\-\int_{0}^{\infty}{\int_{0}^{\Delta x_{\mathrm{n}}/2}{0.9024r_2(\tilde{\nu}_2,5\tilde{\nu}_1)\rho_g\frac{\overline{\epsilon}_{g2}\overline{\epsilon}_{g1}}{\tilde{\nu}_2}dx}dy}-\int_{0}^{\infty}{\int_{0}^{\Delta x_{\mathrm{n}}/2}{3.1043r_2(\tilde{\nu}_1,\tilde{\nu}_1)\rho_g\frac{\overline{\epsilon}_{g1}^2}{\tilde{\nu}_2}dx}dy}.
\end{split}
\label{eq:smallvoidfraction}
\end{equation}

If we define the small bubble gas mass flux as: $Q_{\mathrm{g1}}=\int_{0}^{\infty}{\int_{0}^{\Delta x_{\mathrm{n}}/2}{\rho_g\overline{\epsilon}_{g1}(\overline{w}_{l}+\overline{w}_{\mathrm{rel},1})dx}dy}$ and if we take $Q_{\mathrm{g1}}^{*}$ to be the right hand side of Eq. \ref{eq:smallvoidfraction} we get:

\begin{equation}
    \pder{Q_{\mathrm{g1}}}{z}=Q^{*}_{\mathrm{g1}}.
    \label{eq:smallbubsflux}
\end{equation}

The asterisk denotes the fact that the quantity has extra units "per meter" as compared to its counterpart without an asterisk. 

Integration of the transport equation for the large bubble void fraction, taking into account the symmetry conditions analogous to the small bubble distribution, gives

\begin{equation}
\begin{split}
\pder{}{z}\int_{0}^{\infty}{\int_{0}^{\Delta x_{\mathrm{n}}/2}{\rho_{\mathrm{g}}\overline{\epsilon}_{\mathrm{g2}}(\overline{w}_{\mathrm{l}}+\overline{w}_{\mathrm{rel},2})dx}dy}=-\int_{0}^{\infty}{\int_{0}^{\Delta x_{\mathrm{n}}/2}{Z_2\rho_{\mathrm{g}}\overline{\epsilon}_{\mathrm{g2}}dx}dy}\\+\int_{0}^{\infty}{\int_{0}^{\Delta x_{\mathrm{n}}/2}{0.9024r_2(\tilde{\nu}_2,5\tilde{\nu}_1)\rho_{\mathrm{g}}\frac{\overline{\epsilon}_{\mathrm{g2}}\overline{\epsilon}_{\mathrm{g1}}}{\tilde{\nu}_2}dx}dy}+\int_{0}^{\infty}{\int_{0}^{\Delta x_{\mathrm{n}}/2}{3.1043r_2(\tilde{\nu}_1,\tilde{\nu}_1)\rho_{\mathrm{g}}\frac{\overline{\epsilon}_{\mathrm{g1}}^2}{\tilde{\nu}_2}dx}dy}.
\end{split}
\label{eq:largevoidfraction}
\end{equation}
Note that the right hand side of Eq. \ref{eq:largevoidfraction} is equal to the right hand side of Eq. \ref{eq:smallvoidfraction} with the opposite sign. Eq. \ref{eq:largevoidfraction} can be simplified to

\begin{equation}
    \pder{Q_{\mathrm{g2}}}{z}=Q^{*}_{\mathrm{g2}}=-Q^{*}_{\mathrm{g1}}.
     \label{eq:largebubsflux}
\end{equation}

\subsubsection{Transport equations for the mean bubble volumes}

In addition to the void fraction, transport equations for the mean bubble volumes $\tilde{\nu}_1$ and $\tilde{\nu}_2$ are required. For the small bubbles, this is given by \citep{bohne2020development,lehr2002bubble} 
\begin{equation}
\begin{split}
\pder{}{x}\left(\rho_{\mathrm{g}}\overline{\epsilon}_{\mathrm{g1}}\tilde{\nu}_1 \overline{u}_{\mathrm{l}}\right)+\pder{}{y}\left(\rho_{\mathrm{g}}\overline{\epsilon}_{\mathrm{g1}}\tilde{\nu}_1 \overline{v}_{\mathrm{l}}\right)+\pder{}{z}\left(\rho_{\mathrm{g}}\overline{\epsilon}_{\mathrm{g1}}\tilde{\nu}_1 \overline{w}_{\mathrm{g1}}\right)=-Z_1\tilde{\nu}_1\rho_g\overline{\epsilon}_{\mathrm{g1}}\\+0.3463r_2(\tilde{\nu}_1,\tilde{\nu}_1)\overline{\epsilon}_{\mathrm{g1}}^2\rho_{\mathrm{g}}+Z_2\rho_{\mathrm{g}}\overline{\epsilon}_{\mathrm{g2}}\tilde{\nu}_1-0.9024r_2(\tilde{\nu}_2,5\tilde{\nu}_1)\rho_{\mathrm{g}}\frac{\overline{\epsilon}_{\mathrm{g2}}\overline{\epsilon}_{\mathrm{g1}}}{\tilde{\nu}_2}\tilde{\nu}_1-3.1043r_2(\tilde{\nu}_1,\tilde{\nu}_1)\rho_{\mathrm{g}}\frac{\overline{\epsilon}_{\mathrm{g1}}^2}{\tilde{\nu}_2}\tilde{\nu}_1.
\end{split}
\label{eq:transportsmallbubs}
\end{equation}
Spatial integration of  this equation and applying the symmetry conditions leads to

\begin{equation}
\begin{split}
\pder{}{z}\int_{0}^{\infty}{\int_{0}^{\Delta x_{\mathrm{n}}/2}{\rho_{\mathrm{g}}\overline{\epsilon}_{\mathrm{g1}}\tilde{\nu}_1(\overline{w}_{\mathrm{l}}+\overline{w}_{\mathrm{rel},1})dx}dy}=-\int_{0}^{\infty}{\int_{0}^{\Delta x_{\mathrm{n}}/2}{Z_1\tilde{\nu}_1\rho_g\overline{\epsilon}_{\mathrm{g1}}dx}dy}\\+\int_{0}^{\infty}{\int_{0}^{\Delta x_{\mathrm{n}}/2}{0.3463r_2(\tilde{\nu}_1,\tilde{\nu}_1)\overline{\epsilon}_{\mathrm{g1}}^2\rho_{\mathrm{g}}dx}dy}+\int_{0}^{\infty}{\int_{0}^{\Delta x_{\mathrm{n}}/2}{Z_2\rho_{\mathrm{g}}\overline{\epsilon}_{\mathrm{g2}}\tilde{\nu}_1dx}dy}\\-\int_{0}^{\infty}{\int_{0}^{\Delta x_{\mathrm{n}}/2}{0.9024r_2(\tilde{\nu}_2,5\tilde{\nu}_1)\rho_{\mathrm{g}}\frac{\overline{\epsilon}_{\mathrm{g2}}\overline{\epsilon}_{\mathrm{g1}}}{\tilde{\nu}_2}\tilde{\nu}_1dx}dy}-\int_{0}^{\infty}{\int_{0}^{\Delta x_{\mathrm{n}}/2}{3.1043r_2(\tilde{\nu}_1,\tilde{\nu}_1)\rho_{\mathrm{g}}\frac{\overline{\epsilon}_{\mathrm{g1}}^2}{\tilde{\nu}_2}\tilde{\nu}_1dx}dy}.
\end{split}
\label{eq:smallbubsvol}
\end{equation}
It is noteworthy that the mean volume of the small bubbles also depends on the break up of the small bubbles ($Z_1$) and on the coalescence of small bubbles ($r_2(\tilde{\nu}_1,\tilde{\nu}_1)$), which did not contribute to the transport of the void fraction. Defining $K_{\mathrm{g1}}=\int_{0}^{\infty}{\int_{0}^{\Delta x_{\mathrm{n}}/2}{\rho_{\mathrm{g}}\overline{\epsilon}_{\mathrm{g1}}\tilde{\nu}_1(\overline{w}_{\mathrm{l}}+\overline{w}_{\mathrm{rel},1})dx}dy}$ and abbreviating the right hand side of Eq. \ref{eq:smallbubsvol} as $K^{*}_{\mathrm{g1}}$ we get

\begin{equation}
    \pder{K_{\mathrm{g1}}}{z}=K^{*}_{\mathrm{g}1}.
    \label{eq:smallbubsvolsimple}
\end{equation}
Analogously, the integral form of the transport equation for the large bubble volume is given by
\begin{equation}
\begin{split}
\pder{}{z}\int_{0}^{\infty}{\int_{0}^{\Delta x_{\mathrm{n}}/2}{\rho_{\mathrm{g}}\overline{\epsilon}_{\mathrm{g2}}\tilde{\nu}_2(\overline{w}_{\mathrm{l}}+\overline{w}_{\mathrm{rel},2})dx}dy}=-\int_{0}^{\infty}{\int_{0}^{\Delta x_{\mathrm{n}}/2}{2Z_2\tilde{\nu}_2\rho_{\mathrm{g}}\overline{\epsilon}_{\mathrm{g2}}dx}dy}\\+\int_{0}^{\infty}{\int_{0}^{\Delta x_{\mathrm{n}}/2}{1.8048r_2(\tilde{\nu}_2,5\tilde{\nu}_1)\rho_{\mathrm{g}}\overline{\epsilon}_{\mathrm{g1}}\overline{\epsilon}_{\mathrm{g2}}dx}dy}+\int_{0}^{\infty}{\int_{0}^{\Delta x_{\mathrm{n}}/2}{0.4250r_2(\tilde{\nu}_2,\tilde{\nu}_2)\rho_{\mathrm{g}}\overline{\epsilon}_{\mathrm{g2}}^2dx}dy}\\+\int_{0}^{\infty}{\int_{0}^{\Delta x_{\mathrm{n}}/2}{3.1043r_2(\tilde{\nu}_1,\tilde{\nu}_1)\rho_{\mathrm{g}}\overline{\epsilon}_{\mathrm{g1}}^2dx}dy},
\end{split}
\label{eq:largebubsvol}
\end{equation}
which we can write in short as

\begin{equation}
    \pder{K_{\mathrm{g2}}}{z}=K^{*}_{\mathrm{g}2}.
    \label{eq:largebubsvolsimple}
\end{equation}

\subsection{Initial conditions}
\begin{figure}[!ht]
\centering
\includegraphics[width=0.5\textwidth]{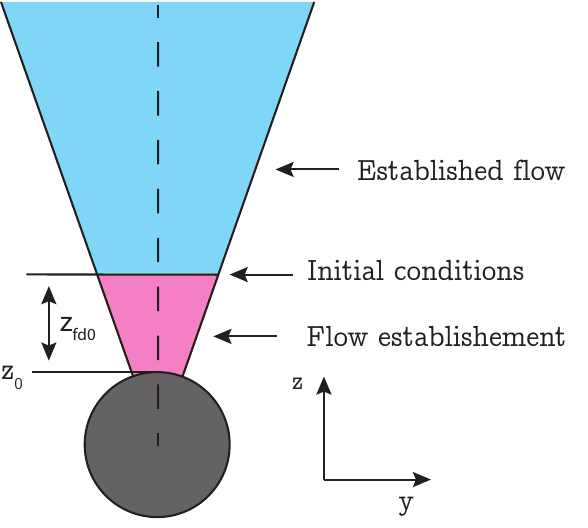}
\caption{Graphical representation of the location where the initial conditions are defined} \label{fig:ICs}
\end{figure}

All initial conditions apply for the zone of established flow (see Figure \ref{fig:ICs}) for which velocity and void fraction distributions can be considered to be Gaussian.  \citet{bohne2020development} took the height of the initial zone of flow establishment as $z_{\mathrm{fd0}}=6.2d_{\mathrm{n}}$ in line with \citet{lee2003turbulent}. The initial height for the integration is thus $z=z_0+z_{\mathrm{fd0}}$.

Since at the initial stage of the bubble curtain we have individual round plumes the initial conditions are the same as given by \citet{bohne2020development} who based it on \citet{milgram1983mean}. The initial mean vertical water velocity can be found through momentum conservation just above the zone of flow establishment:

\begin{equation}
    \overline{w}_{\mathrm{lm0}}=\sqrt{\frac{-2M_{\mathrm{l}0}(2\lambda^2+1)}{\gamma \overline{b}_0 \rho_{\mathrm{l}} \pi (2\overline{\epsilon}_{\mathrm{gm10}} \lambda^2+2\overline{\epsilon}_{\mathrm{gm20}}\lambda^2-2\lambda^2-1)}}.
    \label{eq:ulzm0}
\end{equation}
Here the initial liquid momentum flux $M_{\mathrm{l}0}$ results from the sum of the momentum of the injected air and the buoyancy of the zone of flow establishment according to

\begin{equation}
    M_{\mathrm{l}0}=V_{\mathrm{gn0}}\rho_{\mathrm{g0}} \langle\langle w_{\mathrm{gn0}}\rangle\rangle+2V_{\mathrm{gna}}\frac{p_\mathrm{a}}{p_0}\frac{1}{u_0}(\rho_{\mathrm{l}}-\rho_{\mathrm{a}})gz_{\mathrm{fd0}}.
    \label{eq:inimom}
\end{equation}
Here $V_{\mathrm{gn0}}=V_{\mathrm{gna}}\frac{\rho_{\mathrm{ga}}}{\rho_{\mathrm{g0}}}$ is the initial air flowrate of a single nozzle with $V_{\mathrm{gna}}$ the air flowrate of a single nozzle at ambient pressure ($p_{\mathrm{a}}$), $\rho_{\mathrm{ga}}$ is the gas density at ambient pressure and $\rho_{\mathrm{g0}}$ is the local (i.e. at $z=z_0$) air density at pressure $p_0$. Finally, $\langle\langle w_{\mathrm{gn0}}\rangle\rangle$ denotes the average gas velocity through the nozzle.\\

Based on the conservation of air mass, the initial width of the round plumes can be determined from

\begin{equation}
    \overline{b}_{0}=\sqrt{\frac{V_{\mathrm{g0}}\rho_{\mathrm{g0}}(\lambda^2+1)}{\lambda^2\pi\rho_{\mathrm{g0}}(\lambda^2+1)(\overline{\epsilon}_{\mathrm{gm10}}\overline{w}_{\mathrm{rel},1}+\overline{\epsilon}_{\mathrm{gm20}}\overline{w}_{\mathrm{rel},2})+\lambda^2\overline{w}_{\mathrm{lm0}}\pi\rho_{\mathrm{g0}}(\overline{\epsilon}_{\mathrm{gm10}}+\overline{\epsilon}_{\mathrm{gm20}})}}.
    \label{eq:b0}
\end{equation}
Due to their interdependence, Eq. \ref{eq:ulzm0} and Eq. \ref{eq:b0} must be solved iteratively.

In line with \citet{milgram1983mean} the initial centerpoint void fraction is 

\begin{equation}
    \overline{\epsilon}_{\mathrm{gm0}}=0.5.
\end{equation}
And the initial centerline gas fractions of the large and small bubbles is taken in accordance with \citet{bohne2020development} to be

\begin{equation}
\begin{split}
    \overline{\epsilon}_{\mathrm{gm20}}=0.99\overline{\epsilon}_{\mathrm{gm0}}\\
    \overline{\epsilon}_{\mathrm{gm10}}=\overline{\epsilon}_{\mathrm{gm0}}-\overline{\epsilon}_{\mathrm{gm20}}=0.01\overline{\epsilon}_{\mathrm{gm0}}.
\end{split}
\label{eq:epsini}
\end{equation}
The initial mean large bubble volume can be determined from the so-called `primary bubble radius' ($a_{\mathrm{prim}}$). The primary bubble radius is determined by the bubble pinch-off from a pressurized nozzle due to buoyancy and has been the subject of many studies \citep[e.g.][]{badam2007experimental, gaddis1986bubble, jamialahmadi2001study,kulkarni2005bubble,ramakrishnan1969studies}. \citet{bohne2020development} used the force balance as given by \citet{voit1987calculation} for the primary bubble radius.
In all of the aformentioned studies the primary bubble radius increases with the air flowrate at the nozzle $V_{\mathrm{gn0}}$, and with the exception of the relation by \citet{jamialahmadi2001study} all predictions agree closely. It should be noted, however, that expressions for $a_{\mathrm{prim}}$ have only been validated for flowrates up to $V_{\mathrm{gna}}\approx 0.05\,\mathrm{Ls^{-1}}$, while the flowrate in our measurements ranged between $V_{\mathrm{gna}}=0.03\,\mathrm{Ls^{-1}}$ and $V_{\mathrm{gna}}=0.6\,\mathrm{Ls^{-1}}$.

In lack of a better alternative and for consistency with \citet{bohne2020development}, we adopt the relation of \citet{voit1987calculation} for the primary bubble radius, which is the result of the force balance at the nozzle
\begin{equation}
    a_{\mathrm{prim}}=\frac{1}{2}\left[\frac{6}{\pi} \frac{F_{\eta}+F_{\mathrm{T}}+F_{\sigma}}{\rho_{\mathrm{l}}g}\right]^{1/3}.
\end{equation} 
Where $F_{\eta}=15\eta\frac{V_{\mathrm{gn0}}}{2a_{\mathrm{prim}}}$ is the viscous force with $\eta$ the dynamic viscosity of water. $F_{\mathrm{T}}=1.3\rho_{\mathrm{l}}\left(\frac{V_{\mathrm{gn0}}}{2 a_{\mathrm{prim}}}\right)^2$ is the inertial force and $F_{\sigma}=\pi d_{\mathrm{n}}\sigma$ is the surface tension force. These three forces are compared to the gravitational force.
The peak of the exponential distribution of the large bubbles should be at the primary bubble radius, in order to ensure that the mean large bubble volume is

\begin{equation}
    \tilde{\nu}_{20}=\frac{4}{3}\pi\frac{3}{2} a_{\mathrm{prim}}^3=2\pi a_{\mathrm{prim}}^3.
\end{equation}

The initial small bubble mean volume is one of the free parameters in our model. As discussed in detail in section \ref{subsec:nu12}, our data is best approximated when using a flowrate independent initial condition of 
\begin{equation}
    \tilde{\nu}_{10}=40\,\mathrm{mm^3}.
\end{equation}

\subsection{Solving}

The set of 6 integral equations to be solved (Eqs. \ref{eq:Momentumresultsimp},         \ref{eq:massresult},  \ref{eq:smallbubsflux},  \ref{eq:largebubsflux}, \ref{eq:smallbubsvolsimple}, \ref{eq:largebubsvolsimple}) depends on the 6 unknowns $\mathbf{\Theta}=\left[\overline{w}_{\mathrm{lm}}, \overline{b}, \overline{\epsilon}_{\mathrm{gm1}}, \overline{\epsilon}_{\mathrm{gm2}}, \tilde{\nu}_1, \tilde{\nu}_2 \right]^T$. The full system can be written as

\begin{equation}
     \pder{}{z}\mathbf{\Psi}(\mathbf{\Phi},z)=\boldsymbol{\zeta}(\mathbf{\Phi},z),
    \label{eq:tobesolved}
\end{equation}
where $\mathbf{\Psi}=[M,Q_{\mathrm{l}},Q_{\mathrm{g1}},Q_{\mathrm{g2}},K_{\mathrm{g1}},K_{\mathrm{g2}}]^T$ contains the integral quantities on the left-hand side of the equations and $\boldsymbol{\zeta}=[B,\alpha \overline{w}_{\mathrm{lm}}\rho_{\mathrm{l}}P,Q^{*}_{\mathrm{g1}},Q^{*}_{\mathrm{g2}},K^{*}_{\mathrm{g1}},K^{*}_{\mathrm{g2}}]^T$ contains the right-hand sides. The explicit dependence of $\mathbf{\Psi}$ on $z$ enters due to the variation of the gas density with height. To find the solution, we apply the product rule to yield
\begin{equation}
     \pder{\mathbf{\Theta}}{z}=\pder{\mathbf{\Psi}}{\mathbf{\Theta}}^{-1}\left(\boldsymbol{\zeta}-\pder{\mathbf{\Psi}}{z}\right).
    \label{eq:solved}
\end{equation}
The components of the matrix $\pder{\mathbf{\Psi}}{\mathbf{\Theta}}$ can be calculated analytically for all variables except for $\overline{b}$, for which we estimate the derivative numerically: $\pder{\mathbf{\Psi}}{\overline{b}}\approx \frac{\mathbf{\Psi}(1.01\overline{b})-\mathbf{\Psi}(0.99\overline{b})}{0.02\overline{b}}$. The integrals are evaluated using the trapezoidal rule, where the $y$-direction is cut off at $5\overline{b}$. Consistently, we also limit the summation of nozzles in the $x$-direction in Eqns. \ref{eq:vel2} and \ref{eq:void2} to those that lie within $|x|\leq 5\overline{b}$, such that an increasing number of nozzles is considered as the width of the plume grows.\\
Since bubbles categorized as `small' coalesce to form 'large' bubbles, the mean bubble volume of the large bubbles can not be smaller than the small mean bubble volume. This constraint can be enforced by setting
\begin{equation}
       \pder{\tilde{\nu}_2}{z}= \begin{cases}
      \pder{\tilde{\nu}_2}{z} & \text{if } \tilde{\nu}_2>\tilde{\nu}_1\\
      \pder{\tilde{\nu}_1}{z} & \text{if } \tilde{\nu}_2<\tilde{\nu}_1\, .
    \end{cases}  
\end{equation}
However since this results in a discontinuity it increases the demands on the accuracy of the solver and thus increases computational time. To circumvent this we instead use

\begin{equation}
       \pder{\tilde{\nu}_2}{z}= \begin{cases}
      \mathrm{erf}\left[\left(\frac{\tilde{\nu}_2}{\tilde{\nu}_1}\right)-1\right]\pder{\tilde{\nu}_2}{z} + \mathrm{erfc}\left[\left(\frac{\tilde{\nu}_2}{\tilde{\nu}_1}\right)-1\right]\pder{\tilde{\nu}_1}{z}& \text{if } \tilde{\nu}_2>\tilde{\nu}_1\\
      \pder{\tilde{\nu}_1}{z} & \text{if } \tilde{\nu}_2<\tilde{\nu}_1\, .
    \end{cases}  
\end{equation}
Note that in the derivation by \citet{lehr2002bubble} they even assume the large bubbles to be significantly larger than the small bubbles, so the proposed implementation is a very minimal interpretation of this condition.
\\
With this relaxed discontinuity, the set of ordinary differential equations can be solved quickly on a regular PC and using standard solvers (e.g. the ode45 function in MATLAB). We employed an absolute tolerance of $10^{-6}$ and a relative tolerance of $10^{-3}$. 

\section{Results} \label{sec:results}

In this section, we will present our experimental results and compare to model predictions where appropriate. 

The void fraction profile has been measured at different heights in the bubble curtain using electrical probes. These probes principally measure the contact time of a bubble with the needle tip. 

\begin{figure}[!ht]
\centering
\includegraphics[width=\columnwidth]{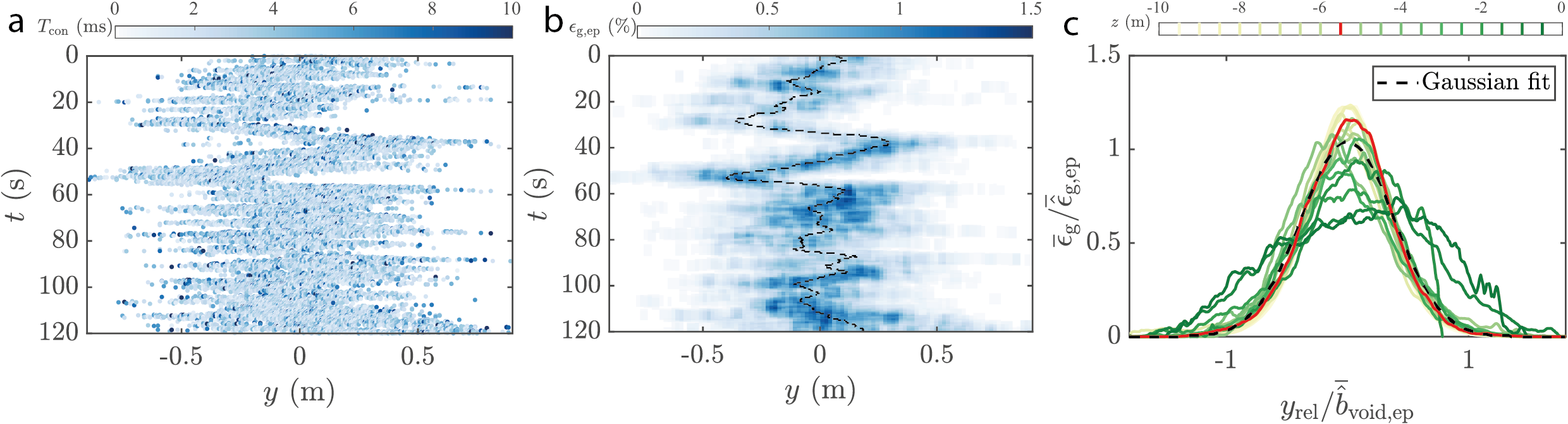}
\caption{Impression of the data processing steps taken. (a) Time-sequence of bubble contact times at $V^*_{\mathrm{ga}} = 6.2\,\mathrm{Lm^{-1}s^{-1}}$ and $z =-5.5\,\mathrm{m}$, (b) void fraction distribution corresponding to the signal in (a), dashed line represents the centerline, (c) normalised void fraction distributions for $V^*_{\mathrm{ga}} = 6.2\,\mathrm{Lm^{-1}s^{-1}}$ at different heights.} \label{fig:hittimevoid}
\end{figure}

In Figure \ref{fig:hittimevoid}a a sample time sequence of contact times \citep[corrected to account for the bubble-needle interaction][]{beelen2023situ} is shown. From the corrected contact time we can obtain the instantaneous void fraction distribution across the bubble curtain (in the $y$-direction) by averaging the binary contact signal (1 in air, 0 otherwise) using a moving average with a window of 5 seconds in time and spatially over 5 needles. This results in Figure \ref{fig:hittimevoid}b for the signal shown in Figure \ref{fig:hittimevoid}a.
To obtain robust values for the amplitude and the width of the void fraction distribution, we employ the top-hat definitions

\begin{equation}
\overline{\hat{\epsilon}}_{\mathrm{g,ep}}=\frac{\overline{E}_2}{\overline{E}_1}
\label{eq:alphahat}
\end{equation} 
and
\begin{equation}
\overline{\hat{b}}_{\mathrm{void,ep}}=\frac{\overline{E}_1}{\overline{\hat{\epsilon}}_{\mathrm{g,ep}}},
\label{eq:what}
\end{equation} 
based on the integrals $\overline{E}_1=\int_{\Lambda} \overline{\epsilon}_{\mathrm{g,ep}} dy$ and $\overline{E}_2=\int_{\Lambda}\overline{\epsilon}_{\mathrm{g,ep}}^2dy$, with $\Lambda$ denoting the length of the measurement rake. The subscript $\mathrm{ep}$ stands for electrical probes. 
We compensate for the slow wandering of the bubble curtain by taking the average profile relative to the centerline of the bubble curtain (dashed line Figure \ref{fig:hittimevoid}b). The resulting profiles normalized by the top hat quantities (Figure \ref{fig:hittimevoid}c) are self-similar and resemble a Gaussian distribution (the fitted Gaussian is described by $\overline{\epsilon}_{\mathrm{g,ep}}/\overline{\hat{\epsilon}}_{\mathrm{g,ep}}=1/(0.38\sqrt{2\pi})\exp{\left(-(y_{\mathrm{rel}}/\overline{\hat{b}}_{\mathrm{void,ep}})^2/0.38^2\right)}$) consistent with the assumption in section \ref{subsec:3.4} up to a depth of about $z = -2\,\mathrm{m}$ at which surface effects start to become relevant. 

\subsection{Entrainment factor}
\label{subsec:entrainres}
\begin{figure}[!ht]
\centering
\includegraphics[width=\columnwidth]{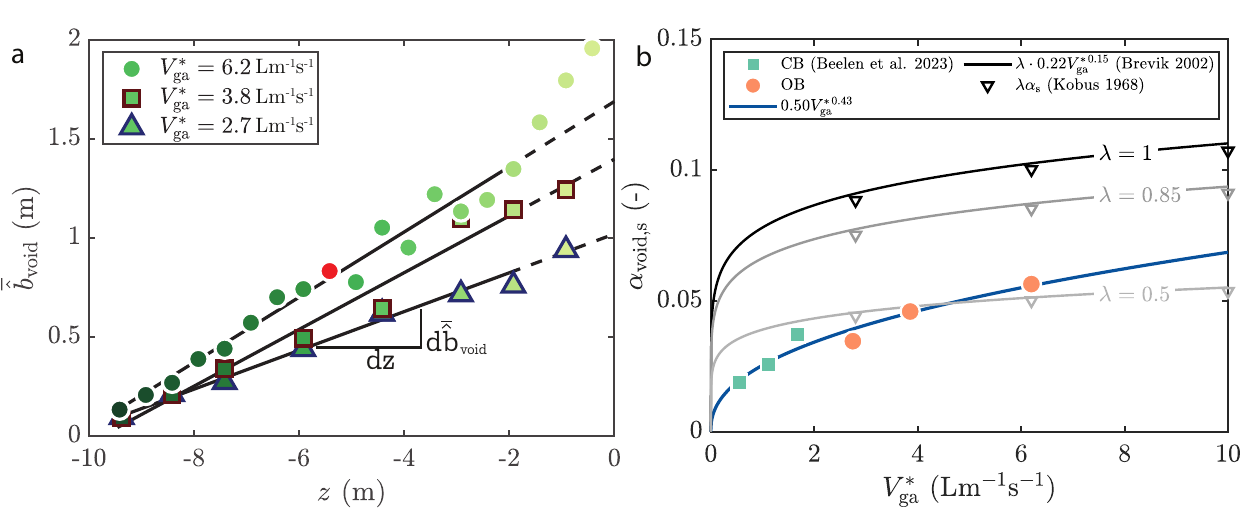}
\caption{a) Measured top-hat width of the void fraction field for different air flowrates. b) Results for $\alpha_{void,s}$ as a function of  $V_{ga}^{*}$. For comparison, data and fit by \citet{kobus1968analysis} and \citet{brevik2002flow} are also included assuming different values of $\lambda$}\label{fig:entrainfactor}
\end{figure}

To characterise the spreading of the plume, we plot $\overline{\hat{b}}_{\mathrm{void}}$ as a function of $z$ at $V_{\mathrm{ga}}^* = 6.2\,\mathrm{Lm^{-1}s^{-1}}$ (corresponding to the data in Figure \ref{fig:hittimevoid}) as well as two other flowrates in Figure \ref{fig:entrainfactor}a. 
Noting that for a Gaussian distribution $\overline{\hat{b}}_{\mathrm{void}}=\sqrt{2\pi}\overline{b}_{\mathrm{void}}$, we can obtain the spreading angle based on the Gaussian half-width of the void fraction distribution from
\begin{equation}
    \alpha_{\mathrm{void,s}} = \frac{1}{2\sqrt{2}} \frac{d \overline{\hat{b}}_{\mathrm{void}}}{d z}.
\end{equation}
We compute values of $\alpha_{\mathrm{void,s}}$ by fitting the slope in the region in which $\hat{b}_{\mathrm{void}}$ grows approximately linearly with $z$ (indicated by the solid lines). The corresponding results are presented in Figure \ref{fig:entrainfactor}b along with complementing data of \citet{beelen2023situ} at lower $V^*_{\mathrm{ga}}$. These data are are well approximated by the empirical fit 
\begin{equation}
    \alpha_{\mathrm{void,s}}=0.5V^{* \, 0.43}_{\mathrm{ga}}.
    \label{eq:entrianrelation}
\end{equation}
For reference, we have also included results by \citet{kobus1968analysis} in the figure. Since they reported 
data for $\alpha_{\mathrm{s}}$, i.e. based on the velocity distribution, a comparison is only possible when assuming a value for $\lambda=\overline{b}_{\mathrm{void}}/\overline{b}$, which leads to $\alpha_s = 1/\lambda \alpha_{\mathrm{void,s}}$. According to \citet{milgram1983mean}, a reasonable range is $0.8<\lambda<0.9$ and therefore we use $\lambda=0.85$. As can be seen in Figure \ref{fig:entrainfactor}b, the data of \citet{kobus1968analysis} and the fit to that data by \citet{brevik2002flow} $\alpha_{\mathrm{s}} = 0.22 V_{\mathrm{ga}}^{*\, 0.15}$ significantly differ from ours at this value and agreement is only observed for an unreasonably low value of $\lambda = 0.5$, hinting that entrainment in bubble plumes additionally depends on other factors such as the nozzle type as observed in \citet{beelen2023situ}. Also the impact of the wandering of the bubble curtain on the properties in \citet{kobus1968analysis} is unknown. In our model, we adopt Eq. \ref{eq:entrianrelation}, since a more complete parameterization, although efforts have been made for confined round plumes (\citep{seol2007particle,lima2018effect}),
is not available yet for unconfined planar plumes.

\subsection{Comparison of integral scales}
We now compare the model predictions to our data. The values of all parameters in the model used for this comparison are documented in Table \ref{tab:table1}. 

\begin{table}[!ht]
    \centering
    \begin{tabular}{|c|c|c|c|}
        Variable & Description & Value & Reference \\\hline
        $c_{\varepsilon}$ & Dissipation constant & $0.024$ & \citet{zhao2014evolution}\\
        $u_{\mathrm{crit}}$ & Minimum coalescence velocity & $0.08\,\mathrm{ms^{-1}}$ & \citet{bohne2020development}\\
        $\gamma$ & Momentum amplification factor & 1 & \citet{bohne2020development}\\
        $\epsilon_{\mathrm{gmax}}$ & Maximum packing fraction air in water & $0.6$ & \citet{millies1999interfacial}\\
        $\eta$ & Dynamic viscosity water & $10^{-3}\,\mathrm{Pa\cdot s}$ & \\
        $\lambda$ & Width ratio void fraction and velocity field & $0.85$ & \citet{milgram1983mean}\\
        $\sigma$ & Surface tension water air & $72\cdot 10^{-3}\,\mathrm{Nm^{-1}}$ & 
        \\
        $\rho_{\mathrm{ga}}$ & Ambient air density & $1.2\,\mathrm{kgm^{-3}}$ & \\
        $\rho_{\mathrm{l}}$ & Water density & $1000\,\mathrm{kgm^{-3}}$ &   
    \end{tabular}
    \caption{Quantities used throughout the results section, if  applicable a reference has been given}
    \label{tab:table1}
\end{table}

\begin{figure}[!ht]
\centering
\includegraphics[width=\textwidth]{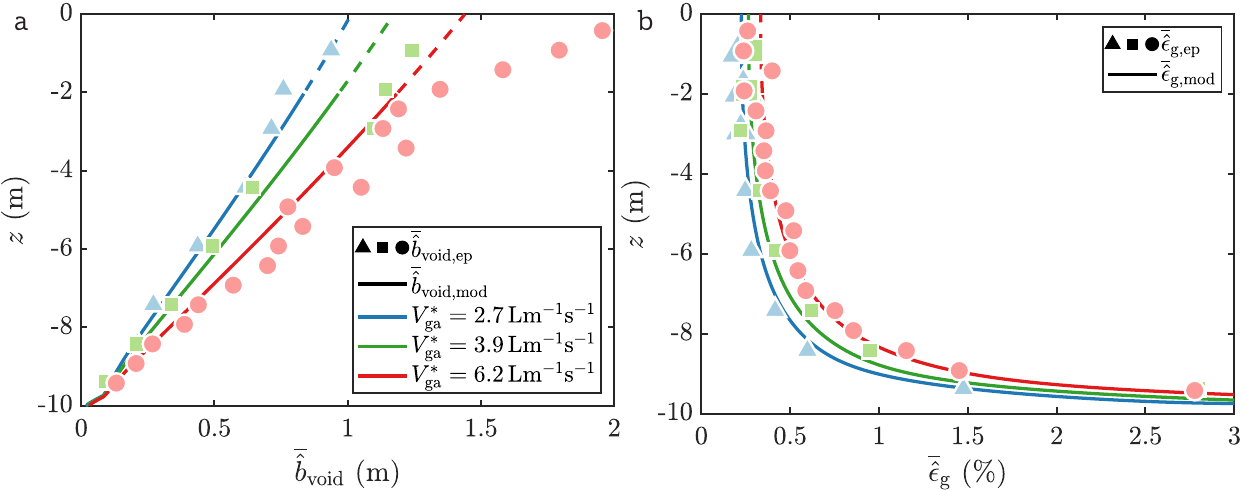}
\caption{a) Top hat width of the model $\overline{\hat{b}}_{\mathrm{void,mod}}$ compared to the measured top hat width $\overline{\hat{b}}_{\mathrm{void,ep}}$ in the OB b) modelled top hat void fraction compared to measured average top hat void fraction}\label{fig:Widthresult}
\end{figure}

Results for the width of the bubble curtains in terms of $\overline{\hat{b}}_{\mathrm{void,mod}}$ and $\overline{\hat{\epsilon}}_{\mathrm{g,ep}}$ are shown in Figure \ref{fig:Widthresult} for the current set of measurements. 
Outside of the recirculation zone (roughly top 20\% according to the relation given in \citet{brevik2002flow}, indicated by dashed lines), the model predictions agree closely with the data. A notable result is that higher gas flowrates predominantly lead to wider bubble curtains while the effective void fraction settles to about $\overline{\hat{\epsilon}}_{\mathrm{g,ep}} \approx 0.5\%$ within the first 2m above the nozzle with only a slight dependence on $V_{\mathrm{ga}}^*$.

\begin{figure}[!ht]
\centering
\includegraphics[width=\textwidth]{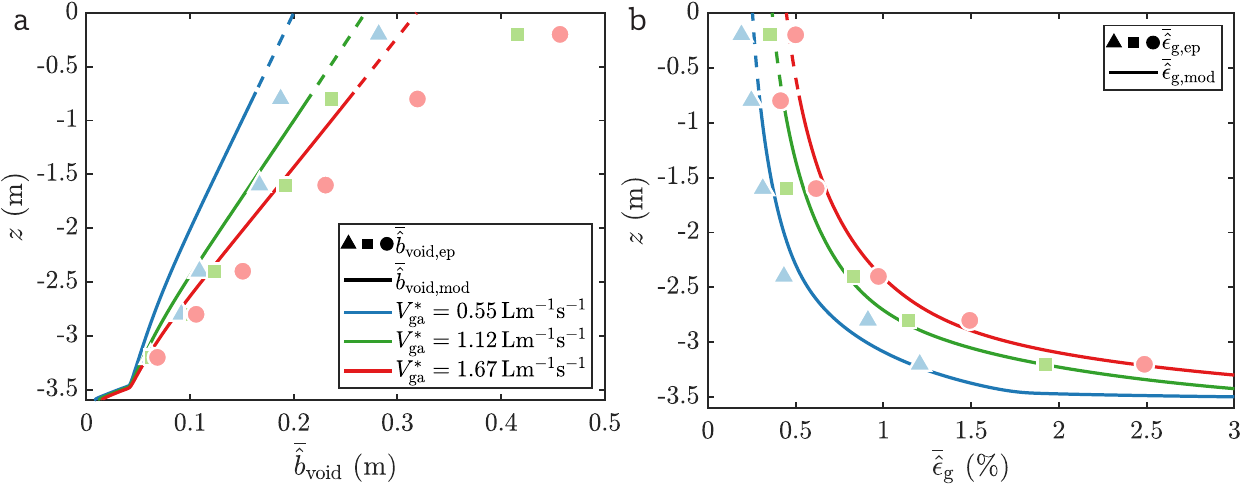}
\caption{Model comparison to the data of \citet{beelen2023situ}. (a) Top hat width of the model $\overline{\hat{b}}_{\mathrm{void,mod}}$ compared to the measured top hat width $\overline{\hat{b}}_{\mathrm{void,ep}}$ in the CB b) modelled top hat void fraction compared to measured average top hat void fraction} \label{fig:VoidOBCB}
\end{figure}

Additionally, we have applied our model to the data presented in \citet{beelen2023situ} and the corresponding results are shown in Figure \ref{fig:VoidOBCB}. In this case, the void fractions are again predicted with very good accuracy, whereas the top-hat width of the model consistently underpredicts the measured values. The reason for this discrepancy is not entirely clear. Potentially, the effective slip velocity assumed in the model is too high at the lower gas flowrates considered here, where collective effects might play less of a role (see section \ref{subsec:slipvel}). It should be noted, however, that adjustments leading to a better match of $\overline{\hat{b}}_{\mathrm{void,mod}}$ most likely lead to worse representations of the results for $\overline{\hat{\epsilon}}_{\mathrm{g,ep}}$.

The transition from round to planar plumes and the accompanying change in the entrainment coefficient leads to a kink in the profiles of $\overline{\hat{b}}_{\mathrm{void,mod}}$ in Figures \ref{fig:Widthresult}a and \ref{fig:VoidOBCB}a. The difference in the spreading renders the width of the bubble curtain slightly dependent on the nozzle spacing, since for larger $\Delta x$ the round plume zone featuring a higher $\alpha_{\mathrm{s}}$ is extended and therefore the 'starting width' of the straight plume can be different. 

\subsection{Comparison of bubble size distributions}\label{subsec:4.3}

\begin{figure}[!ht]
\centering
\includegraphics[width=\textwidth]{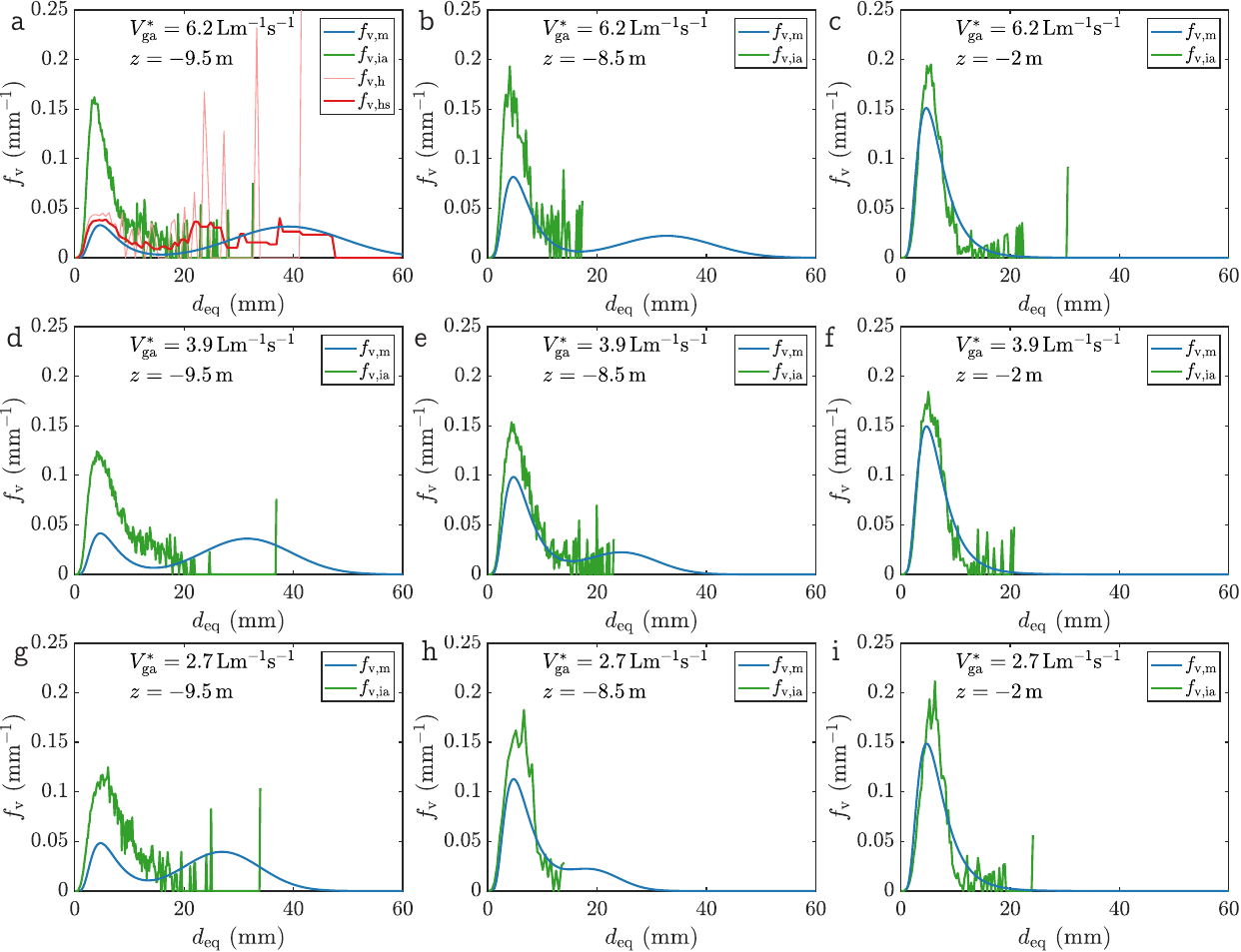}
\caption{Modelled results vs measured results for:
$V^*_{\mathrm{ga}}=6.2\,\mathrm{Lm^{-1}s^{-1}}$ (a-c), $V^*_{\mathrm{ga}}=3.9\,\mathrm{Lm^{-1}s^{-1}}$ (d-f) and $V^*_{\mathrm{ga}}=2.7\,\mathrm{Lm^{-1}s^{-1}}$ (g-i) at different heights of $z= - 9.5m$ (a,d,g), $z= - 8.5m$ (b,e,h) and $z= - 2.0m$ (c,f,i)} \label{fig:Bubsizevsmeas}
\end{figure}

In Figure \ref{fig:Bubsizevsmeas} model predictions for the bubble size distribution are compared to the results from the measurements in the OB at different depths and gas flowrates. To extract the experimental bubble size distributions, we employ the image processing algorithm described in \citet{beelen2023situ} with corresponding results labelled as $f_{\mathrm{v,ia}}$ and shown as green lines in Figure \ref{fig:Bubsizevsmeas}. However, this automated method has limitations at high void fractions and  for very large deformed bubbles. To assess how this affects results close to the nozzle ($z=-9.5\,\mathrm{m}$, Figures \ref{fig:Bubsizevsmeas}a,d,g), where the shortcomings are most prevalent, we additionally processed the measurement at the highest air flowrate $V^*_{\mathrm{ga}}=6.2\,\mathrm{Lm^{-1}s^{-1}}$ by hand. Even though this dataset consists of more than 2000 individual bubbles, weighting by volume results in a spiky signal in this case (see $f_{\mathrm{v,h}}$ in Figure \ref{fig:Bubsizevsmeas}a), which we smooth by applying a moving average with a variable window side of $22/60d_{\mathrm{eq}}$ to yield  $f_{\mathrm{v,hs}}$. Comparing $f_{\mathrm{v,hs}}$ and $f_{\mathrm{v,ia}}$ in Figure \ref{fig:Bubsizevsmeas}a clearly shows that the latter is underestimating the volume contained by the larger bubbles substantially. This is related to the fact that in the algorithm bubbles or clusters of bubbles touching the edge of the image are excluded and larger bubbles are more likely to be part of these structures, as illustrated in appendix \ref{app:appendixA}. While it was too laborious to verify this explicitly, it therefore appears safe to assume that the large-bubble content is similarly underrepresented also in the other cases closer to the nozzle shown in Figure \ref{fig:Bubsizevsmeas}. This is further corroborated by the analysis in appendix \ref{app:appendixA}, where we show that the difference between $f_{\mathrm{v,ia}}$ and the model at smaller value of $d_{eq}$ is related to the normalisation of the distribution. Keeping this in mind and judging by Figure \ref{fig:Bubsizevsmeas}a, the experimentally obtained ($f_{\mathrm{v,ia}}$ or $f_{\mathrm{v,hs}}$) and modelled ($f_{\mathrm{v,m}}$) bubble size distributions are found to be in good agreement. In particular $f_{\mathrm{v,hs}}$ clearly features the bimodal distribution also predicted by the model. Due to breakup of the bubbles, the peak at larger $d_{\mathrm{eq}}$is diminished at  $z = -8.5\,\mathrm{m}$ (Figures \ref{fig:Bubsizevsmeas}b,e,h)  and totally absent at $z = -2\,\mathrm{m}$ (Figures \ref{fig:Bubsizevsmeas},c,f,i).
 
It can also be seen, particularly by comparing Figures \ref{fig:Bubsizevsmeas}c, f  and i, that the measured bubble size distribution away from the nozzle is largely independent of the air flowrate. This holds not only for the range of $V^*_{\mathrm{ga}}$ displayed in Figure \ref{fig:Bubsizevsmeas} but also at the lower flowrates investigated in \citet{beelen2023situ} (not shown).

\section{Discussion}
\label{sec:Discussion}
In this section, we will discuss some aspects and outcomes from the model in more detail. 

\subsection{Void fraction}
\label{subsec:5.1}
\begin{figure}[!ht]
\centering
\includegraphics[width=\textwidth]{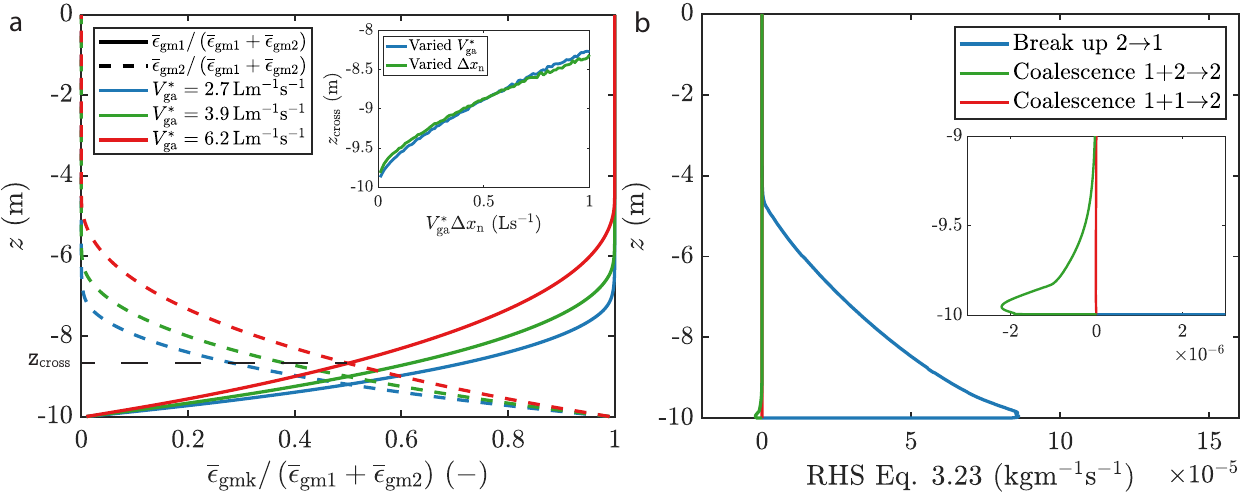}
\caption{a) Relative contribution of the small bubbles (solid lines) and the large bubble to the void fraction (dashed lines) for different air flowrates. The inset gives $z_{\mathrm{cross}}$ as a function of $V_{\mathrm{ga}}^*\Delta x_{\mathrm{n}}$ b) Evaluated terms on the RHS of Eq. \ref{eq:smallvoidfraction} with $V_{\mathrm{ga}}^*=6.2\,\mathrm{Lm^{-1}s^{-1}}$ and $\Delta x_{\mathrm{n}}$} \label{fig:Voidrelative}
\end{figure}

In Figure \ref{fig:Voidrelative}a the relative contribution of small ('1') and large ('2') bubbles to the total void fraction is shown for different air flowrates considered in the experiments. In all cases, the large bubbles initially carry almost all the void volume in line with Eq. \ref{eq:epsini}. However, this changes quickly moving away from the nozzle and after roughly $z=-4\,\mathrm{m}$ the entire gas volume is carried by the small bubbles for all considered air flowrates. 
There is a slight tendency for larger bubbles to persist longer at higher flowrates. We can quantify this by evaluating the cross-over height $z_{\mathrm{cross}}$ of the distributions. As the inset in Figure \ref{fig:Voidrelative}a shows, this quantity depends on the flowrate through a single nozzle $V_{\mathrm{ga}}^*\Delta x_{\mathrm{n}}$, since curves for $z_{\mathrm{cross}}$ at varying $V_{\mathrm{ga}}^*$ and $\Delta x_{\mathrm{n}}$ collapse when plotted as a function of this quantity. 

The shift towards 'small' bubbles is governed by Eq. \ref{eq:smallvoidfraction}, for which the terms on the right hand side  are evaluated in Figure \ref{fig:Voidrelative}b. As this plot shows, the breakup of large into small bubbles (positive here as it increases the amount of small bubbles) is by far dominant while coalescence to form large bubbles is only slightly non-zero very close to the nozzle but remains negligible throughout. Also breakup is strongest close to the nozzle and its relevance declines strongly with increasing $z$.

\subsection{Initial condition for the mean bubble volumes}
\label{subsec:nu12}
\begin{figure}[!ht]
\centering
\includegraphics[width=\textwidth]{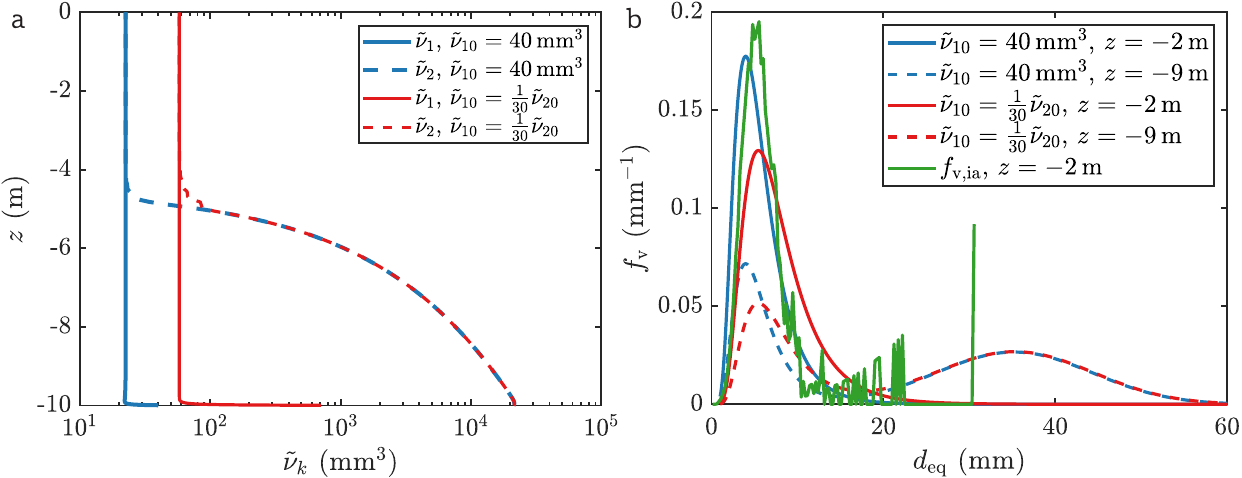}
\caption{a) Mean bubble volumes for different choices for the initial conditions with $V_{\mathrm{ga}}^*=6.2\,\mathrm{Lm^{-1}s^{-1}}$ and $\Delta x_{\mathrm{n}}=0.1\,\mathrm{m}$ b) Resulting bubble size distributions at different heights for both initial conditions compared to data at $z= -2m$} \label{fig:BubsizedistriIC}
\end{figure}

An important difference in our model compared to previous work is the initial condition for the mean bubble volume of the small bubbles. \citet{bohne2020development} set $\tilde{\nu}_{10}$ proportional to the mean volume of the large bubbles according to $\tilde{\nu}_{10}=\frac{1}{30}\tilde{\nu}_{20}$. Since they did not provide a motivation for this choice and given that the implied dependence of $\tilde{\nu}_{10}$ on the air flowrate proved to be inconsistent with our experiments, we instead chose $\tilde{\nu}_{10}=40\,\mathrm{mm^3}$. 
In Figure \ref{fig:BubsizedistriIC}a we explored the impact of the initial small bubble mean volume on the resulting bubble size distribution. To that end we compared our initial condition of $\tilde{\nu}_{10}=40\,\mathrm{mm^3}$ to the initial condition by \citet{bohne2020development}.

As can be seen in Figure \ref{fig:BubsizedistriIC}a, our choice for  $\tilde{\nu}_{10}$ results in significantly smaller bubbles throughout the domain. For both initial conditions the mean small bubble volume decreases close to the nozzle. However, the constants achieved thereafter are different such that the effect of the different initial conditions persists throughout the domain. This is due to the fact that the breakup of small bubbles, the main mechanism by which $\tilde{\nu}_1$ changes (see Eq. \ref{eq:smallbubsvol}), requires high dissipation rates which only exist close to the nozzle.  
The initial condition for $\tilde{\nu}_{20}$ remains unaltered, and the evolution of the mean large bubble volume is similar in the relevant region up to $z = -5\mathrm{m}$, where the large bubble distribution contributes to the void fraction (see Figure \ref{fig:Voidrelative}a). 

The corresponding changes in the volume weighted bubble size distributions can be seen in Figure \ref{fig:BubsizedistriIC}b. 
While the large bubble peak is independent of the initial condition for $\tilde{\nu}_{10}$, the initial condition proposed by \citet{bohne2020development} results in a shift of the small-bubble peak towards larger $d_{\mathrm{eq}}$. The current choice is in better agreement with the data, as the comparison at $z=-2\,\mathrm{m}$ shows, where the large bubbles are no longer relevant.

\subsection{Water velocity}
\begin{figure}[!ht]
\centering
\includegraphics[width=\textwidth]{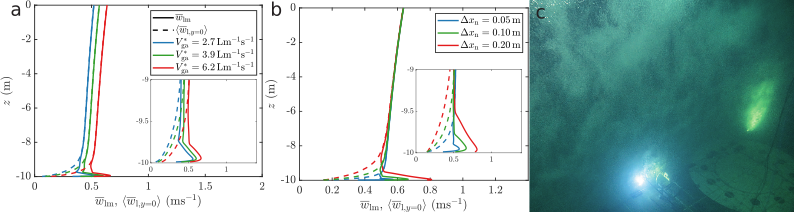}
\caption{The centerpoint velocity (solid line) and the centerline velocity (dashed line) a) for different flowrates with $\Delta x_{\mathrm{n}}=0.1\,\mathrm{m}$ b) for different nozzle spacing with $V^*_{\mathrm{ga}}=6.2\,\mathrm{Lm^{-1}s^{-1}}$. c) Overview image showing the structure of the bubble plume, the grid lines on the bottom of the tank are $2\,\mathrm{m}$ apart} \label{fig:Velocityresults}
\end{figure}

The water velocity has not been measured, therefore we focus on the modelling results. The velocity profile is governed by the average centerpoint velocity $\overline{w}_{\mathrm{lm}}$. However, since the plume transitions from round plumes to a straight plume the velocity will also be presented in terms of the average velocity along the line $y=0$ 
\begin{equation}
    \langle\overline{w}_{\mathrm{l,y=0}}\rangle=\frac{2}{\Delta x_{\mathrm{n}}}\int_0^{\frac{1}{2}\Delta x_{\mathrm{n}}}{\overline{w}_{\mathrm{l}}(x,0)}dx,
\end{equation}
which we call the centerline velocity.\\
In Figure \ref{fig:Velocityresults}a  the centerpoint velocity $\overline{w}_{\mathrm{lm}}$ evaluated for the experimental conditions has a maximum roughly at $z=-9.9\,\mathrm{m}$. For $-9.9<z<-9.7\,\mathrm{m}$ the plumes start to touch and the gradient of the centerpoint velocity clearly changes. From $z\approx -9\,\mathrm{m}$ on wards the centerpoint and centerline velocity are the same indicating that the plume has transitioned to a straight plume. With increasing air flowrate the water velocity in the plume increases. It can also be seen that $\overline{w}_{\mathrm{lm}}$ is not constant as expected for self similar plumes (\citet{carazzo2006route}) but increases with increasing $z$. This is a result of the increase in buoyancy as a consequence of the air expanding with decreasing pressure higher in the water column.

In Figure \ref{fig:Velocityresults}b, we evaluate the effect of changing the nozzle spacing on the flow velocity at constant $V_{\mathrm{ga}}^*$. These results show that the influence of the initial condition quickly decays. Even for the largest spacing for $\Delta x_{\mathrm{n}} = 0.2\,\mathrm{m}$, the effect is only noticeable in the velocities for the first 1m to 2m and eventually the same velocity is reached for all nozzle spacings. 

\subsection{Slip velocity} \label{subsec:slipvel}
Here we discuss our choice for the effective slip velocity, for which we adopted a value of $\overline{w}_{\mathrm{rel}} = 0.8\,\mathrm{\,\mathrm{ms^{-1}}}$, which is significantly higher than the one of $\overline{w}_{\mathrm{rel}} = 0.3\,\mathrm{\,\mathrm{ms^{-1}}}$, which \citet{bohne2020development} used based on the terminal rise velocity of single bubbles. Since we did not measure the water velocity, it is not possible to directly evaluate these different choices based on our own experiments. However, we can obtain an estimate based on the data of \citet{kobus1968analysis}, who measured the fluid velocity distribution along with the integral void fraction $\int_{-\infty}^{\infty}\overline{\epsilon}_{\mathrm{g}}dy$ via the absorption of radioactive radiation. Considering steady-state and accounting for the gas expansion via the ideal gas law we have 
 \begin{equation}
    V^*_{\mathrm{ga}}\frac{p_{\mathrm{a}}}{p}=\int_{-\infty}^{\infty}\overline{\epsilon}_{\mathrm{g}} \overline{w}_{\mathrm{g}}dy.
\end{equation}
Assuming Gaussian straight plume distributions for both vertical velocity and gas fraction, this can be transformed to yield 
\begin{equation}
V^*_{\mathrm{ga}}\frac{p_{\mathrm{a}}}{p}/\int_{-\infty}^{\infty}\overline{\epsilon}_{\mathrm{g}}dy
= \frac{1}{\lambda^2+1}\overline{w}_{\mathrm{lm}}+\overline{w}_{\mathrm{rel}}
\label{eq:weighingvelworkedout}
\end{equation}
Since the left-hand side as well as $\overline{w}_{\mathrm{lm}}$ is known for the measurements of \citet{kobus1968analysis}, we can use this to estimate $\overline{w}_{\mathrm{rel}}$ for their data. Doing so we yields slip velocities in the range of $0.5<\overline{w}_{\mathrm{rel}}<0.6\,\mathrm{ms^{-1}}$ for values of $\lambda$ in the range from 0.8 to 0.9. These estimates significantly exceed the slip velocity of an isolated bubble (see e.g. \citet{haberman1956experimental}) or the values reported for low void fraction confined plumes (\citet{delnoij1999ensemble}). Even higher effective slip velocities that are more in line with the value adopted in our model are reported by \citet{neto2008bubbly}, who find slip velocities of up to $0.7\,\mathrm{ms^{-1}}$ in an unconfined round bubble plume originating from a nozzle. Similarly, \citet{simonnet2007experimental} find even higher slip velocities of up to $0.8\,\mathrm{ms^{-1}}$ albeit in a confined setup with much higher void fractions. 
Taken together, these provide support for the fact that the effective slip of the bubbles exceeds that typically measured for a single bubble. We can confirm that at least indirectly based on our camera images. Our attempts to obtain an estimate for the bubble rise velocity through image tracking failed as the frame rate of $30\,\mathrm{Hz}$ turned out to be too low to capture most bubbles more than once. With a height of the measurement volume of $60\,\mathrm{mm}$, this implies a minimum bubble velocity of $\langle \langle\overline{w}_{\mathrm{g}}\rangle\rangle>\frac{1}{2}0.06\mathrm{m}\times30\mathrm{Hz}=0.9\,\mathrm{ms^{-1}}$. The highest water velocity (centerpoint velocity) is $\approx 0.45\,\mathrm{ms^{-1}}$ according to Figure \ref{fig:Velocityresults}a, such that the resulting lower estimate for the effective slip is $\overline{w}_{\mathrm{rel}}\geq 0.45\,\mathrm{ms^{-1}}$, consistent with the above.

Physically, an enhanced effective slip velocity can be explained through collective effects, i.e. clusters of bubbles locally enhancing the fluid velocity \citep[e.g.][]{ruzicka2000bubbles,krishna1999rise,neto2008bubbly}. Evidence for the existence of such clusters is provided by snapshots from an overview camera (see Figure \ref{fig:Velocityresults}c and the supplementary videos \citep{https://doi.org/10.4121/af9918ff-d1f9-4e96-aece-e4fa22f6bc08.v1}), where cloud like structures in the bubble curtain are clearly visible. The footprint of these structures is also noticeable in the time traces of the void fraction measurements as shown in \citet{beelen2023situ}. The collective behaviour of bubbles also renders a more bubble size independent slip velocity.

Ultimately, the choice of the exact value of $\overline{w}_{\mathrm{rel}} = 0.8\,\mathrm{\,\mathrm{ms^{-1}}}$ is motivated by the agreement with our measured data, in particular for the integral scales shown in section \ref{subsec:entrainres}. It should be noted that for the data of \citet{beelen2023situ} in Figure \ref{fig:VoidOBCB} a slightly lower value for $\overline{w}_{\mathrm{rel}}$ leads to better agreement for the width, whereas the model prediction of the top-hat void fraction becomes significantly worse then. 

\section{Conclusions and Outlook}\label{sec:conclusions}
Our study provides a unique new dataset characterising the bubble distribution of bubble plumes from an array of nozzles. These measurements were performed at considerable scale (10m deep, 31m wide) yet provide details such as the void fraction profile across the planar plume and the bubble size distribution. An important direct outcome is a new parameterisation of the entrainment coefficient in planar bubble plumes. All relevant data will be made publicly available at UT4 for future use. 

We employed this dataset to calibrate and test a hydrodynamical model of the bubble plume. Conceptually, this model is based off the work of \citet{bohne2019modeling,bohne2020development} and \citet{lehr2002bubble} and combines momentum and mass conservation with transport equations for the void fractions and bubble volumes. We extended this approach to seamlessly handle the transition from the initial round plumes to the planar plume configuration further from the model. We further made adaptions to the entrainment relation and the assumed slip velocity in the model and changed the initial condition for the bubble size distribution based on the comparison to our experimental data. With these modifications, we find the model to be in good agreement with the data in terms of the top-hat width and void fraction as well as for the bubble size distribution and its evolution throughout the plume. This fast integral model is an important step in better understanding bubble curtains, and possibly optimizing bubble curtains for their use in sound mitigation of off-shore pile driving. Further progress can be made by combining measurements of the void fraction field such as ours with velocity measurements. In particular this will help to get a better handle on the size ratio $\lambda$ and the effective slip velocity. Future research should then focus on combining this model with acoustic models to predict noise attenuation. 

A final remark concerns the bubble size distribution. Our finding that the bubble size distribution at some distance from the nozzle is independent of the air flowrate, is consistent with previous results of \citet{chmelnizkij2016schlussbericht}. Within the model this observation translates into an initial small bubble volume that is independent of the  air flowrate. However, it is unclear how other environmental effects alter this initial condition, \citet{kulkarni2005bubble} describes some of the possible factors involved. For example salt water is known to suppress coalescence of bubbles (see e.g. \citet{hofmeier1995observations,duignan2021surface,liu2023nanoscale}), and even though coalescence does not play a major role within the model from the initial condition upwards, it should still affect the zone of flow establishment. This could then either affect this initial small bubble mean volume or it could turn out that the bimodal representation of the bubble size distribution no longer holds. Transferring these results to practical applications, especially in seawater, should therefore be done with caution.

\begin{appendix}
\section{Limitations of image algorithm}\label{app:appendixA}
\begin{figure}[!ht]
\centering
\includegraphics[width=\textwidth]{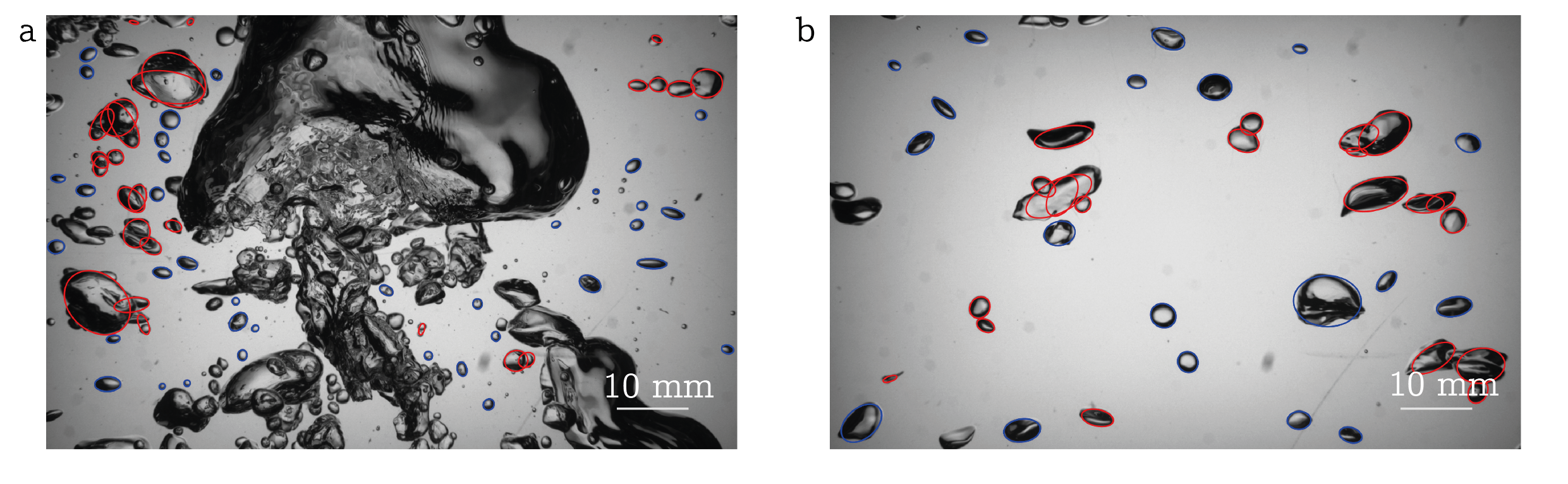}
\caption{a) Typical image close to the nozzle exit b) Typical image far away from the nozzle exit. Bubbles detected by the algorithm are indicated in blue (single bubbles) and red (clusters)}\label{fig:Imagealgorithm}
\end{figure}

Close to the nozzle exit the image algorithm is no longer able to capture the (volume averaged) bubble size distribution. The main reason for this is that as the bubble size approaches that of the field of view, bubbles are increasingly likely to extend beyond the edge of the image. In Figure \ref{fig:Imagealgorithm}a an example of a typical image close to the nozzle is shown including one of the large bubbles. The image algorithm removes all structures touching the edge since it would generally not be able to estimate a bubble size. This means that these bubbles will not be sampled as shown in Figure \ref{fig:Imagealgorithm}. Comparing Figure \ref{fig:Imagealgorithm}a (close to the nozzle) and Figure \ref{fig:Imagealgorithm}b (close to the water surface) we can see that the algorithm will depict a faithful representation of the volume weighted bubble size distribution far away from the nozzle but this is no longer possible close to the nozzle. Apart from not being sampled, large bubbles are typically highly deformed which requires a special treatment as compared to the smaller approximately elliptical bubbles. Additionally, for the cluster deconstruction in our image algorithm to be effective, the overlapping bubbles need to all be part of the outer edge of the cluster, which often is not the case for clusters involving very large bubbles.

\begin{figure}[!ht]
\centering
\includegraphics[width=\textwidth]{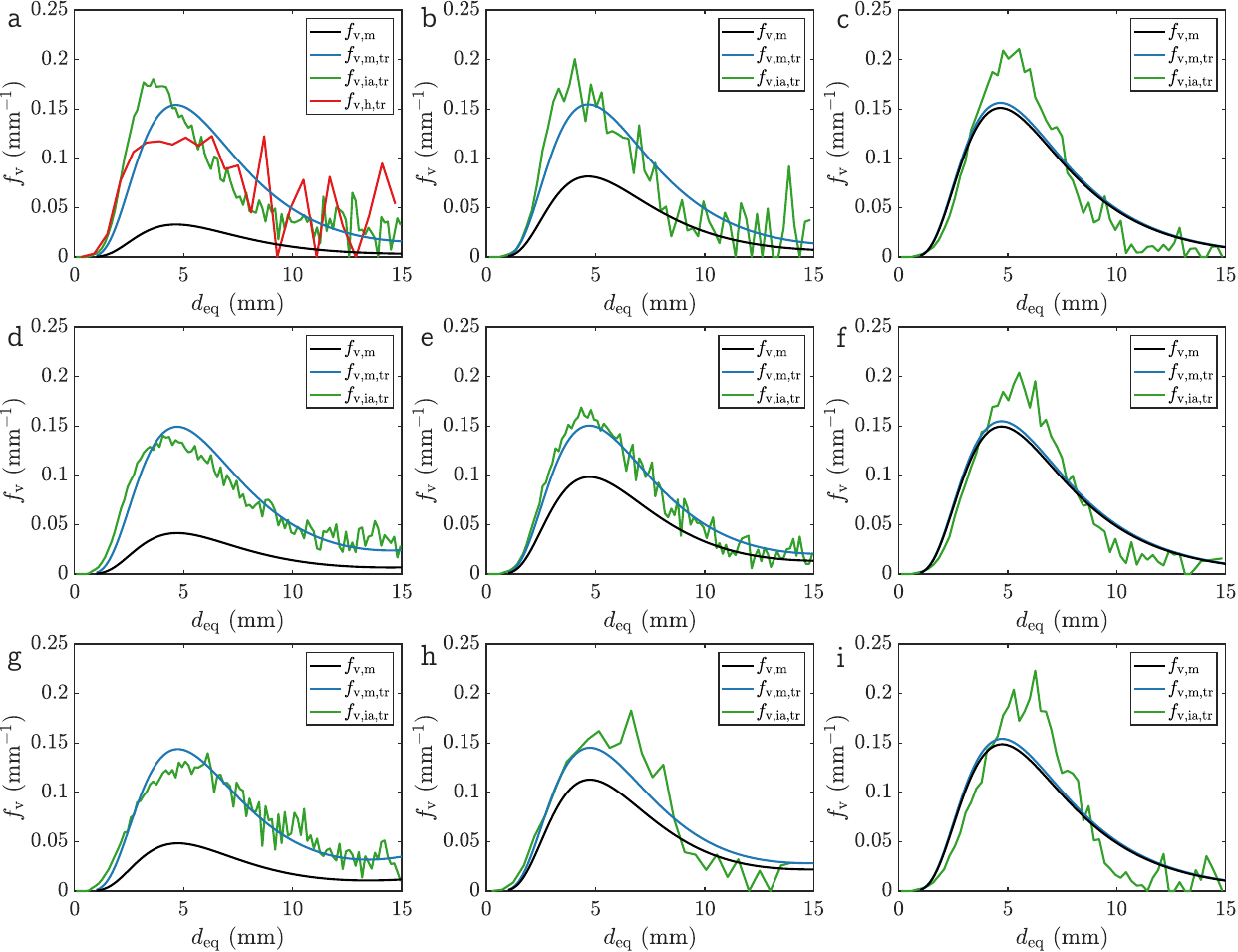}
\caption{Modelled results vs measured results for:
$V^*_{\mathrm{ga}}=6.2\,\mathrm{Lm^{-1}s^{-1}}$ (a-c), $V^*_{\mathrm{ga}}=3.9\,\mathrm{Lm^{-1}s^{-1}}$ (d-f) and $V^*_{\mathrm{ga}}=2.7\,\mathrm{Lm^{-1}s^{-1}}$ (g-i) at different heights of $z= - 9.5m$ (a,d,g), $z= - 8.5m$ (b,e,h) and $z= - 2.0m$ (c,f,i)}\label{fig:fig12truncate}
\end{figure}

To illustrate how the undersampling of large bubbles affects the resulting bubble size distribution, we truncate the bubble size distributions at $d_{\mathrm{eq}}=15\,\mathrm{mm}$ and then renormalise the volume weighted bubble size distribution $f_{\mathrm{v,tr}}$. In Figure \ref{fig:fig12truncate} we show the results for the truncated volume weighted bubble size distributions for all cases considered in Figure \ref{fig:Bubsizevsmeas}, the truncated terms are denoted by the subscript "tr". Alongside the truncated model results we added the original model results for reference. As can be seen, the truncated model, image algorithm result and hand calculated result are very similar if the large bubbles are disregarded. The difference between the model and measured results at small bubble diameters observed at larger depths in figure \ref{fig:Bubsizevsmeas} is therefore only a consequence of the normalization. Since the area under the curve is kept constant, the 'missing' contribution of large bubbles leads to higher values of $f_{v,ia}$ at small $d_{eq}$.

\end{appendix}

\begin{Backmatter}

\paragraph{Acknowledgments}
We gratefully acknowledge the support by MARIN and by the `Bubbles' JIP during the measurements.

\paragraph{Funding Statement}
This publication is part of the project AQUA (with project number P17-07) of the research programme perspectief which is (partly) financed by the Dutch Research Council (NWO). This project is supported by the Netherlands Enterprise Agency (RVO) and TKI Wind op Zee. This project has received funding from the European Research Council (ERC) under the European Union's Horizon 2020 research and innovation programme (grant agreement No. 950111, BU-PACT).

\paragraph{Competing Interests}
None

\paragraph{Data Availability Statement}
The experimental data and the code for the model will be available at UT4 alongside the supplementary videos. \citep{https://doi.org/10.4121/af9918ff-d1f9-4e96-aece-e4fa22f6bc08.v1}. 

\paragraph{Ethical Standards}
The research meets all ethical guidelines, including adherence to the legal requirements of the study country.

\paragraph{Author Contributions}
Simon Beelen: Investigation, Methodology, Software, Formal Analysis, Visualization, Writing original draft\\
Dominik Krug: Conceptualization, Funding acquisition, Supervision, Writing original draft \& review and editing.\\

\bibliography{biblio}

\end{Backmatter}

\end{document}